\documentclass{jpp}
\usepackage{graphicx}
\usepackage{hyperref}
\usepackage[utf8]{inputenc}
\usepackage[T1]{fontenc}
\usepackage{amsmath}
\usepackage{tcolorbox}
\usepackage{amssymb}
\usepackage{xcolor}

\shorttitle{Small-mirror zonal-flow residual}
\shortauthor{E. Rodríguez,  G. G. Plunk}

\title{The zonal-flow residual does not tend to zero in the limit of small mirror ratio}

\author{E. Rodríguez, G. G. Plunk}

\affiliation{Max Planck Institute for Plasma Physics, 17491 Greifswald, Germany}

\begin{document}

\maketitle

\begin{abstract}
The intensity of the turbulence in tokamaks and stellarators depends on its ability to excite and sustain zonal flows.  Insight into this physics may be gained by studying the ``residual'', {\em i.e.} the late-time linear response of the system to an initial perturbation.  We investigate this zonal-flow residual in the limit of a small magnetic mirror ratio, where we find that the typical quadratic approximation to RH \citep{rosenbluth1998poloidal} breaks down. Barely passing particles are in this limit central in determining the resulting level of the residual, which we estimate analytically. The role played by the population with large orbit width provides valuable physical insight into the response of the residual beyond this limit. Applying this result to tokamak, quasi-symmetric and quasi-isodynamic equilibria, using a near-axis approximation, we identify the effect to be more relevant (although small) in the core of quasi-axisymmetric fields, where the residual is smallest. The analysis in the paper also clarifies the relationship between the residual and the geodesic acoustic mode, whose typical theoretical set-ups are similar.  

\end{abstract}

\section{Introduction}
There exists a strong current interest in exploring the space of \textit{stellarators} \citep{spitzer1958stellarator,boozer1998stellarator,helander2014theory}, three-dimensional, toroidal magnetic confinement fields. Optimising such fields in order to achieve plasma confinement and ultimately controlled thermonuclear fusion requires of careful design and shaping of the field for it to present desired physical properties. In guiding this search, it is imperative to have a good understanding of the key physics involved. Given the breadth of the stellarator concept, though, this naturally requires stretching our understanding of physics that are comparatively mature in the simpler case of the axisymmetric \textit{tokamak} \citep{mukhovatov1971,wessonTok}.  
\par
Amongst the critical elements that govern the behaviour of a stellarator, turbulence is a particularly interesting and important one. Understanding the neoclassical behaviour of stellarators has historically captivated much of the focus of research, mainly because of its predominant role in the transport of unoptimised stellarators through the so-called $1/\nu$ regime \citep{galeev1969plasma,stringer1972effect,ho1987,nemov1999,mynick2006}. Progress over the last decades, and especially over the past years \citep{beidler2021demonstration,landreman2022magnetic,goodman2023constructing}, has however brought turbulence to the forefront, and it is now regarded as one of the key elements determining the performance of stellarators.
\par
Zonal flow dynamics are of particular interest in the study of turbulence \citep{diamond2005zonal}, as they are understood to play a key role in regulating turbulence by shearing eddies apart, lowering the overall intensity of turbulent fluctuations. The description of full zonal-flow dynamics is certainly complex, as an essentially non-linear response of the system. However, one may learn some basic information about the ability for a given magnetic equilibrium to sustain such flows by considering the behaviour of the so-called zonal-flow \textit{residual} \citep{rosenbluth1998poloidal, xiao2006short, sugama2006collisionless, monreal2016residual}. The residual is the long-time remnant of an initial radially varying perturbation of the electrostatic potential. The prevalence of a large such remnant is, at least sometimes, indicative of the system's capacity to sustain zonal dynamics in a turbulent state \citep{watanabe2008reduction,xanthopoulos2011zonal}.  The calculation of the residual thus serves as a reasonable starting point for the assessment of zonal flows in a given magnetic equilibrium. The main theoretical understanding of the residual behaviour was pioneered by \cite{rosenbluth1998poloidal}, and subsequently refined and extended by others \citep{xiao2006short,sugama2006collisionless, monreal2016residual,plunk2024residual}, including in the electromagnetic context \citep{catto2017electromagnetic}.
\par
The level of the residual depends strongly on the size of the orbit-width, $\delta$, of the particles in the field, that is, the magnitude of the particle deviation from flux surfaces as they move along field lines. The dependence is so strong that, in a typical scenario \citep{rosenbluth1998poloidal}, it is the trapped particles (whose orbit widths are largest) that contribute most to the residual. The larger the orbit widths, the lower the residual levels, as the shielding from these becomes more effective \citep{rosenbluth1998poloidal,xiao2006short}. In fact, it is conventionally argued that in the limit of $B$ becoming flat (small mirror ratio), the large trapped particle orbits cause the residual to vanish. Of course it is also in this limit that there are also no trapped particles left in the problem, somewhat complicating the asymptotic analysis.  
\par
In this paper we revisit the theoretical question of the zonal-flow residual in this limit.  An assessment is presented in Section~\ref{sec:residual_small_mir}, where we also draw connections to the standard framework of geodesic-acoustic-modes \citep{conway2021geodesic}. We learn that barely passing particles play the dominant role in determining the final finite value of the residual in the small mirror ratio limit. This large-orbit-width part of the population behaves, we argue, as if non-omnigeneous, as far as the residual is concerned. We find support for these claims numerically through linear gyrokinetic simulations. We close the discussion in Section~\ref{sec:field_survey} with an assessment of the relevance of this effect on tokamaks and omnigeneous stellarators, which appears to be limited.

\section{Residual calculation in the small mirror ratio limit} \label{sec:residual_small_mir}
\subsection{Brief derivation of the residual}
Let us start our discussion on the zonal-flow residual by calculating it in its most typical of set-ups. We follow closely the work of \cite{rosenbluth1998poloidal, xiao2006short, monreal2016residual, plunk2024residual}, but include a brief derivation for completeness and as a way of introduction of notation. 
\par
By residual, which we denote $\phi(\infty)$, we mean the surface averaged collisionless electrostatic potential in the long time limit. To describe it, we take the linearised, electrostatic gyrokinetic equation as starting point \citep{CHT,connor1980stability},
\begin{equation}
    \left(\frac{\partial}{\partial t} + i\tilde{\omega}_d + v_\parallel\frac{\partial}{\partial\ell} \right)g = \frac{q}{T}F_0 J_0 \frac{\partial}{\partial t}\phi, \label{eqn:GK}
\end{equation}
written in the ballooning formalism with the variation perpendicular to the field line described by $\mathbf{k}_\perp=k_\psi\nabla\psi$. Here $\psi$ is the flux surface label (the toroidal flux over $2\pi$), so that the electrostatic potential perturbation $\phi$ has a main strong off-surface variation, which is the reason why there is no diamagnetic term in Eq.~(\ref{eqn:GK}), $\omega_\star=0$. Other symbols have their usual meaning: $F_0$ is the background Maxwellian distribution, $J_0=J_0(x_\perp\sqrt{2b})$ the Bessel function of the first kind representing Larmor radius effects and $b=(k_\psi|\nabla\psi|\rho)^2/2$ the Larmor radius parameter, with $\rho=v_T/\Omega$, $v_{T}=\sqrt{2T/m}$ and $\Omega=q\bar{B}/m$ (at this point we are considering a general species of mass $m$, charge $q$ and temperature $T$). The drift frequency $\tilde{\omega}_d=\omega_d (v/v_T)^2(1-\lambda B/2)$ and $\omega_d=\mathbf{v}_D\cdot\mathbf{k}_\perp=v_T \rho \bar{B} k_\psi{\pmb\kappa}\times\mathbf{B}\cdot\nabla\psi/B^2$, with $\bar{B}$ a reference field, ${\pmb \kappa}$ the curvature of the field and the drift is considered in the low $\beta$ limit. The velocity space variables are $\lambda=\mu/\mathcal{E}$ and particle velocity $v=\sqrt{2\mathcal{E}/m}$, where $\mu$ is the first adiabatic invariant and $\mathcal{E}$ the particle energy. The parallel velocity can then be written as $v_\parallel=\sigma v\sqrt{1-\lambda B}$, where $\sigma$ is the sign of $v_\parallel$.
\par
Equation~(\ref{eqn:GK}) is then a partial differential equation in time $t$ and the arc length along the field line $\ell$, for the electrostatic potential $\phi$ and the non-adiabatic part of the distribution function, $g$, with a dependence on the velocity space variables $\{\sigma,v,\lambda\}$. Performing a Laplace transform in time \citep[Theorem~2.7]{schiff2013laplace} yields
\begin{equation}
    \left(\omega - \tilde{\omega}_d + iv_\parallel\frac{\partial}{\partial\ell} \right)\hat{g} = \frac{q}{T}F_0 J_0 \omega\hat{\phi}+i\delta\!F(0), \label{eqn:GK_lap}
\end{equation}
where $\delta\!F(0)\stackrel{\cdot}{=}g(0)-(q/T)J_0F_0\phi(0)$ can be interpreted as the initial perturbation of the system, and we are using the hats to indicate the Laplace transform.
\par
To eliminate the explicit $\ell$ dependence that the curvature, $\tilde{\omega}_d$, brings into the equation, we shall define the orbit width $\delta$,
\begin{equation}
    v_\parallel\frac{\partial}{\partial\ell}\delta=\tilde{\omega}_d - \overline{\tilde{\omega}_d} \label{eqn:def_delta}
\end{equation}
so that we may write,
\begin{equation}
    \left(iv_\parallel\frac{\partial}{\partial\ell}-\overline{\tilde{\omega}_d}+\omega\right)\hat{h}=\frac{q}{T}F_0\omega J_0\hat{\phi} e^{i\delta}+ie^{i\delta}\delta\!F(0), \label{eqn:GK_in_h}
\end{equation}
and $\hat{h}=\hat{g} e^{i\delta}$. The function $\delta$ describes the off-surface displacement of particles (in $\psi$) as a function of $\ell$, for each particle identified by its velocity space labels. The overline notation indicates the bounce average,
\begin{equation}
    \overline{f}=\begin{cases}
    \begin{aligned}
        &\frac{1}{\tau_b}\frac{1}{v}\int_\mathrm{b}  \frac{1}{\sqrt{1-\lambda B}}\sum_\sigma f\,\mathrm{d}\ell, \\
        &\lim_{L\rightarrow\infty}\frac{1}{\tau_t}\frac{1}{v}\int_\mathrm{p}\frac{f\,\mathrm{d}\ell}{\sqrt{1-\lambda B}}.
    \end{aligned}
    \end{cases}\label{eqn:bounce_averages}
\end{equation}
The first expression applies to trapped particles, where the integral is taken between the left and right bounce points and summed over both directions ($\sigma$) of the particle's motion. The normalisation factor is the bounce time, $\tau_b$, defined following $\overline{1}=1$.  For passing particles, the integral is taken over the whole flux surface (i.e. the infinite extent of the field line explicitly indicated by the limit), and normalised by the transit time, $\tau_t$. 
\par
When $\overline{\tilde{\omega}_d}=0$, Eq.~(\ref{eqn:GK_in_h}) simplifies. This corresponds to the physical interpretation of particles having no net off-surface drift. This is the defining property of \textit{omnigeneity} \citep{Hall,Cary1997,Helander_2009,landreman2012}, which we shall assume to hold throughout this work. For a treatment of the non-omnigeneous problem see \cite{helander2011oscillations, monreal2016residual}. 
\par
Because we are interested in the behaviour at large time scales, we expand in $\omega/\omega_t\sim\epsilon_t$, applying $\hat{h}=\hat{h}^{(0)}+\hat{h}^{(1)}+\dots$ and $\hat{\phi}=\hat{\phi}^{(0)}+\hat{\phi}^{(1)}+\dots$, and considering Eq.~(\ref{eqn:GK_in_h}) order by order,
\begin{subequations}
    \begin{align}
        iv_\parallel\frac{\partial}{\partial\ell}\hat{h}^{(0)} &\approx 0, \label{eqn:GK_in_h_order_0} \\
        iv_\parallel\frac{\partial}{\partial\ell}\hat{h}^{(1)}+\omega \hat{h}^{(0)} &\approx \frac{q}{T}\omega F_0J_0\hat{\phi}^{(0)}e^{i\delta}+ie^{i\delta}\delta\!F(0), \label{eqn:GK_in_h_order_1} \\
        & \vdots \nonumber 
    \end{align}
\end{subequations}
From Eq.~(\ref{eqn:GK_in_h_order_0}) it follows that,
\begin{equation}
    \hat{h}^{(0)}=\overline{\hat{h}^{(0)}}.
\end{equation}
Thus, bounce averaging Eq.~(\ref{eqn:GK_in_h_order_1}), and assuming that $\hat{\phi}^{(0)}$ is $\ell$-independent, we may write down the leading order expression for $\hat{g}^{(0)}$,
\begin{equation}
    \hat{g}^{(0)}=\frac{q}{T}F_0 e^{-i\delta}\overline{J_0 e^{i\delta}}\hat{\phi}^{(0)}+\frac{i}{\omega}e^{-i\delta}\overline{\delta\!F(0)e^{i\delta}}.
\end{equation}
With this expression for $\hat{g}$, we may then apply the quasineutrality condition \citep{connor1980stability} summing over ions and electrons. Explicitly, and summing over electrons and ions (subscripts $e$ and $i$ respectively)
\begin{equation}
    \sum_{e,i}\int\mathrm{d}^3\mathbf{v}J_0 \hat{g}=n\frac{q_i}{T_i}(1+\tau)\hat{\phi},
\end{equation}
where $\tau=T_i/ZT_e$ and $Z = -q_i/q_e$, then yields
\begin{equation}
    \hat{\phi}^{(0)}\approx\frac{1}{n}\left\langle \int\mathrm{d}^3\mathbf{v}J_0e^{-i\delta}\overline{J_0e^{i\delta}}F_0 \right\rangle_\psi \hat{\phi}^{(0)}+\frac{i}{\omega}\frac{1}{n}\left\langle \int\mathrm{d}^3\mathbf{v}J_0e^{-i\delta}\overline{\delta\!F(0)e^{i\delta}} \right\rangle_\psi.
\end{equation}
Here $\langle\dots\rangle_\psi$ denotes a flux surface average \citep{helander2014theory}, and we have taken the limit of $m_e\ll m_i$, so that the limit of a negligible electron Larmor radius and electron banana width may be taken; this is equivalent to an adiabatic electron response $\phi-\langle\phi\rangle_\psi$, making the final form of the residual independent of electrons.
\par
By inverse Laplace transforming this latest expression \citep[Theorem~2.36]{schiff2013laplace}, we obtain,
\begin{equation}
    \phi(\infty)=\frac{\frac{1}{n}\left\langle \int\mathrm{d}^3\mathbf{v}J_0e^{-i\delta}\overline{\delta\!F(0)e^{i\delta}} \right\rangle_\psi}{1-\frac{1}{n}\left\langle \int\mathrm{d}^3\mathbf{v}J_0e^{-i\delta}\overline{J_0e^{i\delta}}F_0 \right\rangle_\psi}.
\end{equation}
To finalise the calculation of the residual, we must consider some initial perturbation of the ion population. Following \cite{rosenbluth1998poloidal, monreal2016residual}, we perturb the density of the ions with $\delta\!F(0)=(\delta n/n)J_0 F_0$, a perturbed Maxwellian, sidestepping the issue of detailed initial-condition dependence of the residual, especially important at shorter wavelengths \citep{monreal2016residual}. Applying quasineutrality at $t=0$, the density perturbation may be directly related to the perturbed electrostatic potential $\phi(0)$. Assuming that $b$ is independent of $\ell$ for simplicity, $\delta n/n=\phi(0)(1-\Gamma_0)/\Gamma_0$ where $\Gamma_0=e^{-b}I_0(b)$ and $I_0$ is the Bessel function of the first kind. Therefore, the expression for the residual at long times is,
\begin{equation}
    \frac{\phi(\infty)}{\phi(0)}\approx\frac{1-\Gamma_0}{\Gamma_0}\frac{\frac{1}{n}\left\langle \int\mathrm{d}^3\mathbf{v}J_0e^{-i\delta}\overline{J_0 e^{i\delta}} F_0 \right\rangle_\psi}{1-\frac{1}{n}\left\langle \int\mathrm{d}^3\mathbf{v}J_0e^{-i\delta}\overline{J_0 e^{i\delta}} F_0 \right\rangle_\psi}. \label{eqn:general_RH_expression}
\end{equation}

\subsection{Finite orbit width} \label{sec:orbit_width}
In order to proceed with the evaluation of Eq.~(\ref{eqn:general_RH_expression}) we first need to study the orbits of our particles, namely $\delta$. These will depend critically on both $B(\ell)$ (which controls the time spent by particles along different segments of the field-line), and the normal curvature $\omega_d$ (that determines the off-surface velocity). Although in an actual equilibrium field these functions are connected to each other, it is formally convenient to set this equilibrium connection aside, and treat them as largely independent quantities in the context of a single flux tube.
\par
Despite this independence, it is important to respect some minimal properties. First, for the choice of functions to appropriately represent the behaviour in an omnigeneous field, they should prevent diverging particle orbits. We prevent this ill-behaviour by ensuring that the critical points of $B(\ell)$ match points of zero radial drift; that is, $\omega_d(\ell)=0$ wherever $\mathrm{d}B(\ell)/\mathrm{d}\ell=0$. This property is known as \textit{pseudosymmetry} \citep{mikhailov2002,skovoroda2005}, and is necessary to represent an omnigeneous field. However, it is not sufficient. In addition, we must impose that all the orbits $\delta$ are closed; that is, that they come back to the same $\psi$ at bounce points, or for passing particles, after a period.
\par
With this, we may write explicitly $\delta$ integrating Eq.~(\ref{eqn:def_delta}), as
\begin{equation}
    \delta = \sigma\frac{v}{v_T}\int_{\bar{\ell}_0}^{\bar{\ell}}\frac{1-\lambda B/2}{\sqrt{1-\lambda B}}\frac{\omega_d(\bar{\ell}')}{\omega_t}\mathrm{d}\bar{\ell}' \label{eqn:delta_psi}
\end{equation}
where we have introduced a normalised length scale $\bar{\ell}$ and an associated transit frequency $\omega_t=v_T/L$, with $L$ some reference length scale. The integral is defined so that $\delta(\bar{\ell}_0)=0$, where $\bar{\ell}_0$ corresponds to bounce points for trapped particles, and the point $B=B_\mathrm{max}$ for passing ones to guarantee continuity across the trapped-passing boundary.\footnote{Note that by virtue of omnigeneity it does not matter which point of maximum $B$ or bounce point (left or right) along the field line we choose, because $\delta=0$ at all of these by virtue of omnigeneity. This property of omnigeneous fields is very important, and it allows us to treat each well along the field line independently from every other. This is so because there is no accumulation of radial displacement of passing particles across maxima. Thus, the considerations that the paper presents for a single well could be extended to multiple ommnigeneous wells, treating each separately, and summing their contributions when considering flux surface averages, as needed in Eq.~(\ref{eqn:general_RH_expression}).}
\par
The regularising role of pseudosymmetry at critical points of $B(\ell)$, where it avoids diverging behaviour, can be seen directly from Eq.~(\ref{eqn:delta_psi}). This allows us to rewrite $\delta$ in a form that avoids the explicit $1/\sqrt{\cdot}$ divergence using integration by parts,
\begin{equation}
    \delta = -\sigma \frac{v}{v_T}\left[\frac{B^2}{\partial_{\bar{\ell}} B}\frac{\omega_d(\bar{\ell})}{\omega_t}\frac{\sqrt{1-\lambda B}}{B}\right]_{\bar{\ell}_0}^{\bar{\ell}}+\sigma \frac{v}{v_T} \int_{\bar{\ell}_0}^{\bar{\ell}}\frac{\sqrt{1-\lambda B}}{B}\partial_{\bar{\ell}'}\left(\frac{B^2 }{\partial_{\bar{\ell}'} B}\frac{\omega_d(\bar{\ell}')}{\omega_t}\right)\mathrm{d}{\bar{\ell}}'. \label{eqn:delta_ibp}
\end{equation}
This integrated form of the equation is also useful to numerically compute $\delta$ near bounce points. 
\par
These expressions are so far quite general, and we shall now specialise to a simple representative system. In particular, we assume to have a single unique magnetic well along the field line\footnote{Along any fieldline of an omnigeneous field, every time a maximum of $B$ is crossed, one falls into a new magnetic well. In the case of a tokamak, all those wells are identical by virtue of axisymmetry, and thus the consideration of a single unique well is sufficient. Other optimised configurations, though, lack this exact symmetry, which requires some additional interpretation. Some of this is discussed in Section~\ref{sec:field_survey}.
}, described simply by $B=\bar{B}\left(1-\Delta\cos\pi\bar{\ell}\right)$ and $\omega_d=\omega_d \sin\pi\bar{\ell}$, where the domain is taken to be $\bar{\ell}\in[-1,1]$. Thus the scale $L$ can be interpreted as the connection length in the problem, or the half-width of the well, $\Delta$ the mirror ratio and $\omega_d$ the drift. This particular choice is convenient in two ways: first, because the choice $\omega_d=c\partial_\ell B$, with $c$ some proportioonality constant, simplifies Eq.~(\ref{eqn:delta_ibp}) and conveniently guarantees the closure of particle orbits; and second, because many of the integrals that ensue may be carried out exactly for such simple analytic functions. Of course, deforming these geometric functions away from these forms (in particular, breaking the parity in $\bar{\ell}$) will directly affect the orbit shape $\delta$ and ultimately the residual, but this model nonetheless includes the essential ingredients. 

\begin{figure}
    \centering
    \includegraphics[width = 0.9\textwidth]{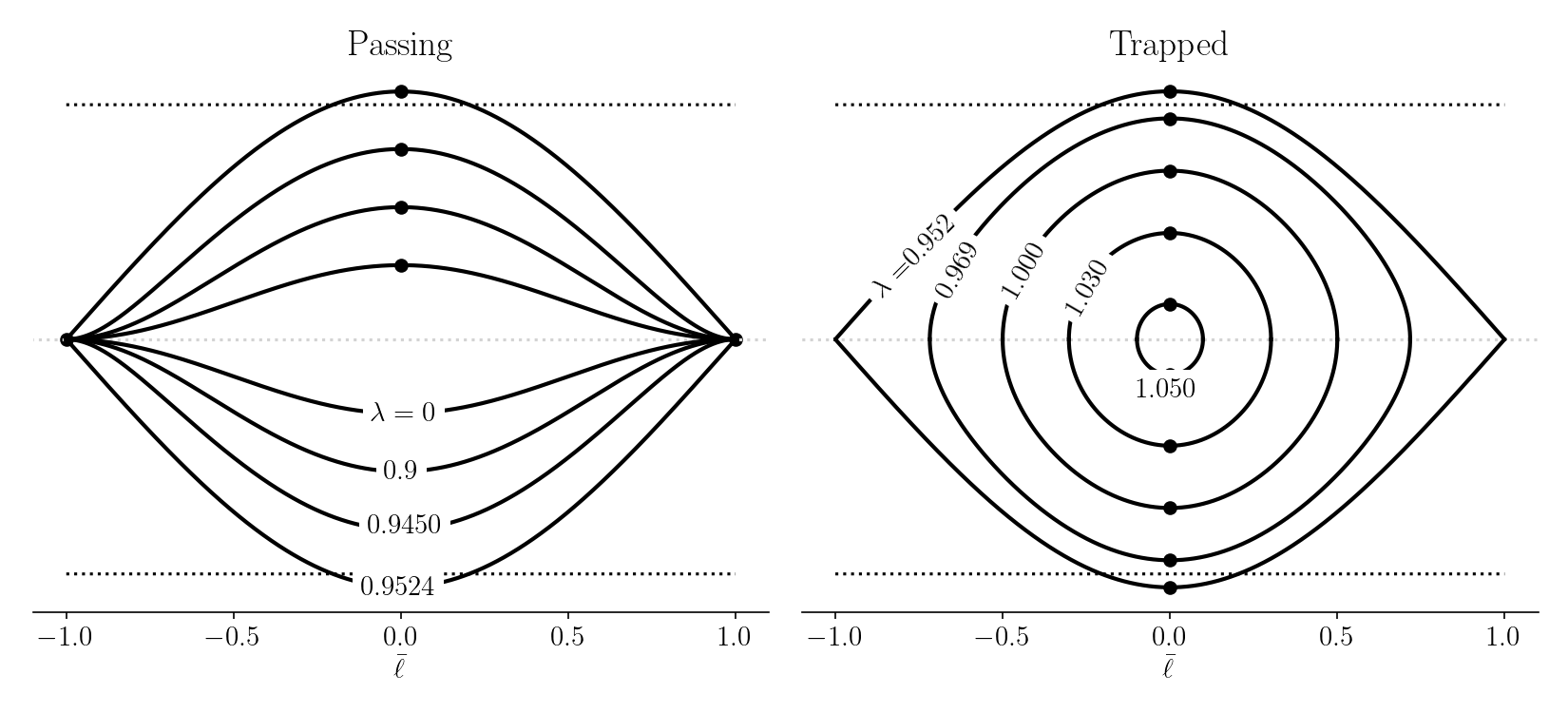}
    \caption{\textbf{Example of passing and trapped orbits.} Numerical examples of trapped and passing orbits for different values of $\lambda$ for the model field considered in the paper. The plots were generated for $\Delta=0.05$. The dotted line on top and bottom correspond to the $\delta(0)$ estimate in Eq.~(\ref{eqn:trap_delta_0}) (grey line simply indicates the reference $\delta=0$ level). Critical points are marked with solid points.}
    \label{fig:particle_orbits}
\end{figure}

\subsubsection{Passing particles} \label{sec:passing_particles}
Let us start our description of the passing particle orbits by considering their maximum deviation off the flux surface, {\em i.e.} their orbit widths $\delta|_{\bar{\ell} = 0} = \delta(0)$. By passing particles we refer to the portion of velocity space with $\lambda\in(0,1/B_\mathrm{max})$, which we may also label with the convenient shifted variable $\hat{\lambda}=1/(1+\Delta)-\bar{B}\lambda$. In this case $\hat{\lambda}=0$ represents the trapped-passing boundary, and $\hat{\lambda}=\bar{B}/B_\mathrm{max}$ is approached for the passing particles far from the trapped-passing boundary, which we will refer to as {\it strongly} passing. It is convenient to introduce yet an additional label for passing particles, namely $\kappa=2\lambda\bar{B}\Delta/[1-\lambda\bar{B}(1-\Delta)]$, which is bounded $\kappa\in(0,1)$ and denotes barely passing particles by $\kappa=1$ and strongly passing by $\kappa=0$.
\par
For the model field considered, $\delta(0)$ may be evaluated exactly in terms of $\lambda$, $\Delta$ and other parameters.
However, it is more insightful to consider some relevant asymptotic limits. In the limit of a small mirror ratio $\Delta$, the passing population is naturally separated into three different regimes, where we may write,
\begin{equation}
    \delta_\mathrm{pass}(0)\approx -\sigma \frac{v}{v_T}\frac{\omega_d}{\pi\omega_t}\times\begin{cases}
                \begin{aligned}
                    &\sqrt{\frac{2}{\Delta}}  &\quad \text{if } \hat{\lambda}\ll\Delta, \\
                    &\frac{1}{\sqrt{\hat{\lambda}}}  &\quad \text{if } \Delta\ll\hat{\lambda}\ll1, \\
                    &\frac{1}{\sqrt{\hat{\lambda}}}+\sqrt{\hat{\lambda}}  &\quad \text{if } \hat{\lambda}\gg\Delta.
                \end{aligned}
           \end{cases}\label{eqn:orbit_width_pass}
\end{equation}
The orbits are widest within a layer of width $\Delta$ near the trapped-passing boundary, where all barely passing particles have large, almost identical orbits that scale like $\sim1/\sqrt{\Delta}$. This is a consequence of particles moving slowly along the field line by an amount $v_\parallel\sim\sqrt{1-\lambda B}\sim\sqrt{\Delta}$. Thus, there always exists a sufficiently small mirror ratio able to slow down \textit{barely passing} particles enough so as for them to have a sizeable orbit width; this is true even for a small radial drift $\epsilon \equiv \omega_d/\pi\omega_t \ll 1$.
\par
We estimate the size of the $v$-space layer that includes particles with a sizeable orbit width (\textit{i.e.} $|\delta_\mathrm{pass}(0)|>1$) in the limit of $\epsilon\ll1$ by taking the behaviour of a typical thermal particle $v/v_T\sim1$ as reference in Eq.~(\ref{eqn:orbit_width_pass}), so that
\begin{equation}
    \hat{\lambda}<\epsilon^2=\left(\frac{\omega_d}{\pi\omega_t}\right)^2. \label{eqn:layer_width}
\end{equation}
Such a layer can only exist if the mirror ratio is sufficiently small,  
\begin{equation}
    \Delta/\epsilon^2\ll1. \label{eq:RH-break-down-limit}
\end{equation}
Not satisfying this mirror ratio ordering restores the standard view of passing particles having small orbit widths (as in the quadratic approximation of the residual in \cite{rosenbluth1998poloidal}). The small mirror ratio ordering alongside the $\epsilon\ll1$ assumption are henceforth assumed.

\subsubsection{Trapped particles} \label{sec:trapped_particles}
The procedure above may be repeated for trapped particles. Defining a trapped particle label $\bar{\kappa}=1/\kappa=[1/(\lambda \bar{B})-(1-\Delta)]/2\Delta$, deeply trapped particles are denoted by $\bar{\kappa}=0$ and barely trapped ones by $\bar{\kappa}=1$. The orbit width may then be written as,
\begin{equation}
    \delta_\mathrm{trap}(0) \approx -\sigma \frac{v}{v_T}\epsilon\sqrt{\frac{2\bar{\kappa}}{\Delta}}, \label{eqn:trap_delta_0}
\end{equation}
assuming $\Delta\ll1$. Unlike passing particles, the majority of trapped particles have a significant orbit width (in the $\sim1/\sqrt{\Delta}$ sense), except for a minute fraction near the bottom of the well which barely moves away from that point. This fraction may be estimated to be
\begin{equation}
    \bar{\kappa}<\frac{\Delta}{\epsilon^2},
\end{equation}
which we have already assumed small. 
\par

\subsection{Evaluating the residual for small mirror ratio} \label{sec:derivation_small_mirror_residual}
\begin{figure}
    \centering
    \includegraphics[width=0.3\textwidth]{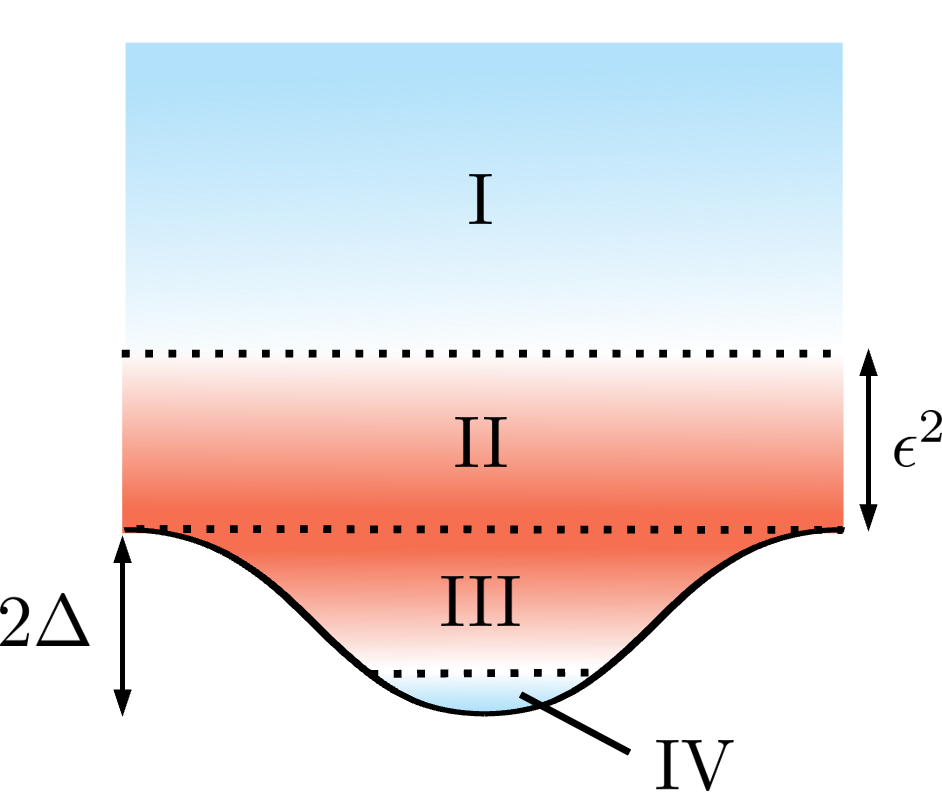}
    \caption{\textbf{Separation of particles into groups.} The diagram depicts the separation of the particle population into four different groups (I to IV). Groups I and IV (light blue) represent the population with a small orbit width, while II and III (light red) correspond to large ones. The diagram is a schematic with the vertical representing $1/\lambda$, the horizontal $\bar{\ell}$ and the black line representing the magnetic well $B(\bar{\ell})$.} 
    \label{fig:particle_groups}
\end{figure}
In the limit of a small mirror ratio, we have learned from the analysis of the orbits that the particle population may be divided into four different groups. Each of these groups is characterised by having a large or small $\delta$, and thus a different contribution to Eq.~(\ref{eqn:general_RH_expression}). We refer to each of these groups by Roman numerals I to IV, starting from strongly passing particles (see Figure~\ref{fig:particle_groups}).
\par
To proceed with the residual integral, let us assume for simplicity the finite-Larmor quantity $b$ to be small. This is compatible with $\epsilon$ being small (note that $\epsilon\propto k_\perp\rho_i$). With this, we may write the integral in the denominator of the residual, Eq.~(\ref{eqn:general_RH_expression}), 
\begin{equation}
    1-\frac{1}{n}\left\langle \int\mathrm{d}^3\mathbf{v}J_0e^{-i\delta}\overline{J_0 e^{i\delta}} F_0 \right\rangle_\psi \approx b-\frac{1}{n}\left\langle \int\mathrm{d}^3\mathbf{v}\left(e^{-i\delta}\overline{e^{i\delta}}-1\right)F_0 \right\rangle_\psi, \label{eqn:separate_flr_res_integral}
\end{equation}
where we used,
\begin{equation}
    \frac{1}{n}\left\langle\int \mathrm{d}^3\mathbf{v}J_0^2 F_0 \right\rangle_\psi=\frac{1}{2}e^{-b}I_0(b),
\end{equation}
in the small $b$ limit and the velocity space integrals include all groups. The integral remaining in Eq.~(\ref{eqn:separate_flr_res_integral}) has been simplified by dropping finite-Larmor radius corrections. For groups I and IV for which $\delta$ is small, retaining $b$ would give an even smaller $O(\delta^2b)$ correction, which we drop. For groups II and III, the correction would also be small in the sense $O(b\sqrt{\Delta})$, under the assumption of small $\Delta$.
\par
Now separating the integral left in Eq.~(\ref{eqn:separate_flr_res_integral}) into the different group contributions,
\begin{multline}
    I=\frac{1}{n}\left\langle \int\mathrm{d}^3\mathbf{v}\left(e^{-i\delta}\overline{e^{i\delta}}-1\right)F_0 \right\rangle_\psi = \sum_{\mathrm{I,~IV}}\frac{1}{n}\left\langle \int\mathrm{d}^3\mathbf{v}\left(\overline{\delta}^2-\overline{\delta^2}\right)F_0 \right\rangle_\psi+ \\
    +\sum_{\mathrm{II,~III}}\frac{1}{n}\left\langle \int\mathrm{d}^3\mathbf{v}\left(e^{-i\delta}\overline{e^{i\delta}}-1\right)F_0 \right\rangle_\psi. \label{eqn:integral_I}
\end{multline}
This separation enables us to exploit the smallness or largeness of $\delta$ accordingly. The smallness of the orbit width for groups I and IV has already been exploited to write the leading order contribution in powers of $\delta$ in the first term of the right-hand side of Eq.~(\ref{eqn:integral_I}). This contribution should be familiar, as it has the quadratic form in which the Rosenbluth-Hinton residual is customarily written \citep{rosenbluth1998poloidal,xiao2006short,plunk2024residual}. We set this part of the calculation aside for now, and focus on the new contributions by groups II and III.

\subsubsection{Contribution from barely passing particles (group II)}
Let us continue our analysis by looking at barely passing particles in group II (see Fig.~\ref{fig:particle_groups}), and their contribution to Eq.~(\ref{eqn:integral_I}),
\begin{equation}
    I_\mathrm{II}=\underbrace{\frac{1}{n}\left\langle \int_\mathrm{II} \mathrm{d}^3\mathbf{v}e^{-i\delta}\overline{e^{i\delta}} F_0 \right\rangle_\psi}_{\textcircled{1}} - \underbrace{\frac{1}{n}\left\langle \int_\mathrm{II} \mathrm{d}^3\mathbf{v}F_0 \right\rangle_\psi}_{\textcircled{2}}. \label{eqn:def_1_2}
\end{equation}
First consider \textcircled{1}, and rewrite it following \cite{xiao2006short} as,
\begin{equation}
    \textcircled{1} = \frac{1}{n}\left\langle \int_\mathrm{II} \mathrm{d}^3\mathbf{v}\left(\overline{\cos\delta}^2+\overline{\sin\delta}^2\right) F_0 \right\rangle_\psi, \label{eqn:I_I_part_1}
\end{equation}
where we have dropped terms odd in $v_\parallel$, annihilated by the integral over velocity space. Note that, although tempting, $\overline{\sin \delta}$ is generally nonzero according to our convention for the bounce average in Eq.~(\ref{eqn:bounce_averages}), where each direction of the passing particles is treated separately. 
\par
To continue with the calculation, we need to evaluate $\overline{\cos\delta}$ explicitly, exploiting that within group II, the function $\delta$ has a large amplitude. As a result, we expect the cosine of $\delta$ to oscillate quickly along $\bar{\ell}$ resulting in an almost exact cancellation. The non-zero contribution may be estimated through the well-known stationary phase approximation \citep[Sec.~6.5]{bender2013advanced},
\begin{equation}
    \overline{\cos\delta}=\frac{1}{\tau_t\omega_t}\frac{v_T}{v}\Re\left\{\int_{-1}^1\frac{e^{i\delta}}{\sqrt{1-\lambda B}}\mathrm{d}\bar{\ell}\right\}\approx \frac{1}{\tau_t\omega_t}\frac{v_T}{v}\sum_i\sqrt{\frac{2\pi}{|\delta''(\ell_i)|}}\frac{\cos[\delta(\ell_i)-\pi/4]}{\sqrt{1-\lambda B(\ell_i)}},
\end{equation}
where the sum is over the turning points of $\delta$ in $\bar{\ell}\in[0,1]$. Using the details of $\delta$ developed in Sec.~\ref{sec:passing_particles} and Appendix~\ref{app:orbit_width}, 
\begin{equation}
    \overline{\cos\delta}\approx\frac{2}{\tau_t\omega_t}\left(\frac{v_T}{v}\right)^{3/2}\sqrt{\frac{\omega_t}{\omega_d}}\left[(4\hat{\lambda})^{-1/4}+\frac{1}{(2\Delta+\hat{\lambda})^{1/4}}\cos\left(\frac{v}{v_T}\frac{\epsilon}{\sqrt{\Delta/2+\hat{\lambda}}}-\frac{\pi}{4}\right)\right].
\end{equation}
The first term inside the square brackets comes from the edge contribution, and the second from the point of maximum excursion. 
\par
Now that we have $\overline{\cos\delta}$ we must integrate over velocity space, Eq.~(\ref{eqn:I_I_part_1}). To do so we introduce the velocity space measure in the $\{v,\lambda,\sigma\}$ coordinate system (already summed over $\sigma$ to give a factor of 2) \citep[Sec.~4.4]{hazeltine2003plasma},
\begin{equation}
    \mathrm{d}^3\mathbf{v}\rightarrow\frac{2\pi B}{\sqrt{1-\lambda B}}v^2\mathrm{d}v\mathrm{d}\lambda.
\end{equation}
and noting that by definition any bounced averaged quantity is $\bar{\ell}$-independent, write for any function $f$ in our single well, 
\begin{equation}
    \left\langle\int_\mathrm{II}\mathrm{d}^3\mathbf{v}\bar{f}\right\rangle_\psi=\pi\bar{B}\int_{v=0}^\infty \int_\mathrm{II}  v^2 \frac{v}{v_T}\tau_t\omega_t \bar{f}\mathrm{d}v\mathrm{d}\lambda, \label{eqn:flux_average}
\end{equation}
correct to leading order in $\Delta$.
\par
The simplifying assumption of a $v$-independent boundary layer in Eq.~(\ref{eqn:layer_width}) allows us to explicitly carry out the integral over $v$ first. Noting the that with the ordering $\epsilon^2/\Delta\gg1$ (large $A$), 
\begin{subequations}
    \begin{align}
        &\int_0^\infty ve^{-v^2}\cos^2\left(Av-\frac{\pi}{4}\right)\mathrm{d}v\approx \frac{1}{4}, \\
        &\int_0^\infty ve^{-v^2}\mathrm{d} v=\frac{1}{2},
    \end{align}
\end{subequations}
we find using the explicit form of the Maxwellian $F_0$,
\begin{equation}
    \textcircled{1}\approx\frac{2}{\sqrt{\pi}}\frac{1}{\omega_d}\int_0^{\epsilon^2}\frac{1}{\hat{\tau}_t}\left(\frac{1}{\sqrt{\hat{\lambda}}}+\frac{1}{\sqrt{2\Delta+\hat{\lambda}}}\right)\mathrm{d}\hat{\lambda},  \label{eqn:contribution_1}
\end{equation}
where $\hat{\tau}_t=\tau_t(v/v_T)$ is a function of $\lambda$. In this form of \textcircled{1} we have already included the contribution from $\overline{\sin\delta}$, which can be easily shown to be equivalent to that of the $\overline{\cos\delta}$. To carry out the integral over $\hat{\lambda}$ we change variables to $\kappa$, defined in Sec.~\ref{sec:passing_particles}. The integration domain becomes $\kappa\in[2\Delta/\epsilon^2,1]$, with an integral measure
\begin{equation}
    \frac{\mathrm{d}\kappa}{\mathrm{d}\lambda}=2\Delta\bar{B}\left(1+\frac{1-\Delta}{2\Delta}\kappa\right)^2.
\end{equation}
The contribution from the edges of the orbit (the first term in Eq.~(\ref{eqn:contribution_1})) can be shown to be small upon integration over $\kappa$ in the limit of small $\Delta$. All that is left is the contribution from the point of maximal excursion, which can be approximated assuming $K(\kappa)\approx \pi/2$,
\begin{equation}
    \textcircled{1}\approx\frac{\epsilon}{\pi^{3/2}}.
\end{equation}
\par
This concludes the calculation of \textcircled{1}, but \textcircled{2} remains to be found. This contribution corresponds to finding the fraction of phase space occupied by the barely passing particles in group II. Using Eq.~(\ref{eqn:flux_average}) and the definition of region II, the integrals over $\kappa$ and $v$ yield,
\begin{equation}
    \textcircled{2}\approx \epsilon.
\end{equation}
Altogether,
\begin{equation}
    I_\mathrm{II}\approx\frac{\epsilon}{\pi^{3/2}}(1-\pi^{3/2})\approx -0.26\frac{\omega_d}{\omega_t},
\end{equation}
yielding an overall negative contribution linear in $k_\psi\rho_i$. 

\subsubsection{Contribution from the bulk of trapped particles (group III)}
A similar approach to that for the barely passing particles may be directly applied to the trapped particles that constitute group III. Given the similarities of the calculation we shall be less explicit here. 
\par
The evaluation of the integral starts once again by separating the integral $I_\mathrm{III}$ into two parts, \textcircled{1} and \textcircled{2}, like in Eq.~(\ref{eqn:def_1_2}). In the calculation of \textcircled{1}, and unlike for passing particles, we only need to consider the  $\overline{\cos\delta}$ term, as $\overline{\sin\delta}=0$ upon summing over both particle directions, Eq.~(\ref{eqn:bounce_averages}). The $\overline{\cos\delta}$ term may be computed much like in the previous section, employing the stationary phase approach. In this case, the only turning point of $\delta_\mathrm{trap}$ is at the centre of the domain, $\bar{\ell}=0$. With that, using the expressions for $\delta_\mathrm{trap}$ introduced in Sec.~\ref{sec:trapped_particles} and Appendix~\ref{app:orbit_width}, and performing the integral over $v$ first, 
\begin{equation}
    \textcircled{1}\approx\frac{1}{\pi^{3/2}}\frac{\Delta}{\epsilon},    
\end{equation}
which is a small contribution that vanishes in the limit of $\Delta\rightarrow0$. The velocity space volume occupied by the bulk of trapped particles, \textcircled{2}, is of course also small in the limit of a small mirror ratio, $\textcircled{2}\sim \sqrt{\Delta}$. Thus, the contribution to the residual from the trapped population in group III is small in the limit of $\Delta\rightarrow0$.

\subsubsection{Final form of the residual}
Gathering the pieces of the calculation above, the integral in Eq.~(\ref{eqn:integral_I}) evaluates to,
\begin{equation}
    I\approx-0.26\frac{\omega_d}{\omega_t}, \label{eqn:I_integral_result}
\end{equation}
in the limit of $\Delta\ll\epsilon^2\ll1$. The latter is particularly important to argue that the contribution from the particles of groups I and IV is subsidiary in this limit. We do not need to compute it explicitly to argue that it scales like $\epsilon^2$, and thus is one order $\epsilon$ higher than the contribution from barely passing particles. Therefore, we may drop those contributions in writing the result in Eq.~(\ref{eqn:I_integral_result}).
\par
We may now write the expression for the residual itself, going back to Eq.~(\ref{eqn:general_RH_expression}) using the definition of $I$ in Eq.~(\ref{eqn:integral_I}),
\begin{equation}
    \frac{\phi(\infty)}{\phi(0)}\approx\frac{1}{1+0.26\frac{\omega_d}{b\omega_t}}, \label{eqn:residual_0p26}
\end{equation}
which in the limit of $b\ll\omega_d$, say for very long radial wavelengths, can be expressed as
\begin{equation}
    \frac{\phi(\infty)}{\phi(0)}\approx 1.92~k_\perp\rho_i\left(\frac{k_\perp\rho_i}{\omega_d/\omega_t}\right). \label{eqn:anal_lay_res}
\end{equation}

\section{Analysis of the residual in the small mirror ratio limit}
The preceding analysis demonstrates that in the limit of a small mirror ratio there remains a finite residual in the problem. Barely passing particles near the passing-trapped boundary dominate the behaviour of the residual in this limit. This is a result of a narrow $\lambda$-space layer of width $\epsilon^2$ having sufficiently slow parallel velocities so that their orbits are wide. The result is a partial shielding of the potential. 
Their orbit width is so large, though, that their shielding is not as efficient as it may be at smaller $\delta$, and thus the residual is larger than one would a priori expect. 
\par
There are two important actors that determine the final value of the residual in this limit: (i) the width of the layer, and (ii) the shape of the orbit. Both of these may be identified directly in the derivation of the residual above. The residual will be larger the smaller the layer is, as the shielding population decreases. The shorter the time that the particles spend near the point of maximal excursion, the larger the residual will also be; orbit shapes that are flat near that point are detrimental to the residual. 
\par
\begin{figure}
    \centering
    \includegraphics[width = \textwidth]{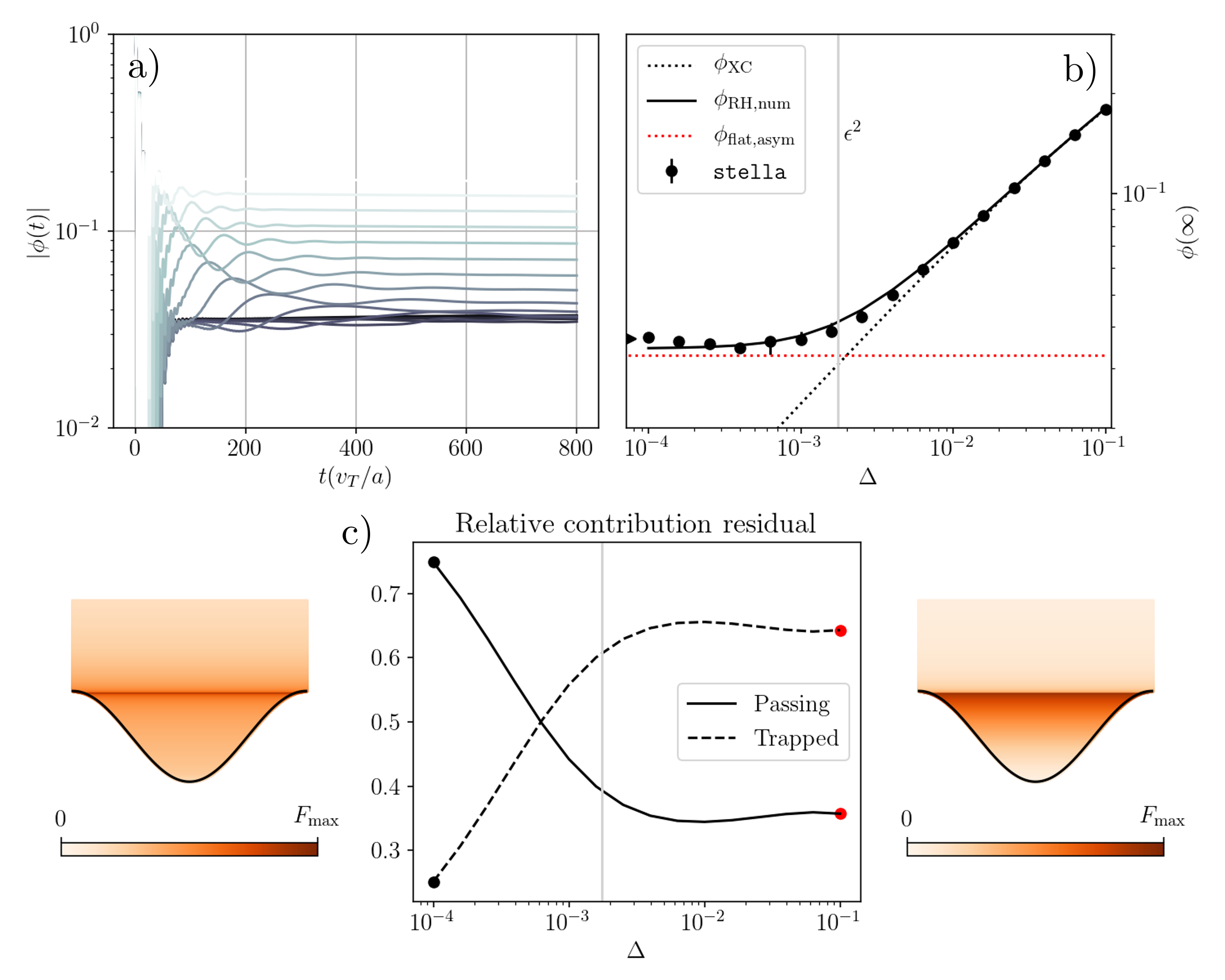}
    \caption{\textbf{Example of residual as a function of mirror ratio.} The plots present (a) the time evolution of the average electrostatic potential for different mirror-ratios simulated with the gyrokinetic code \texttt{stella}, (b) comparison of residual from the gyrokinetic code \texttt{stella} and numerical evaluation of Eq.~(\ref{eqn:general_RH_expression}), and (c) relative contribution to the residual by passing/trapped population, and by each $\lambda$. The simulation for (a) and (b) is based on the cyclone-base-case with $|\mathbf{B}|$ modified, leaving the curvature drift unchanged. The color code in (a) corresponds to the different mirror ratios on the right plot, from lower (darker) to larger (brighter) values of $\Delta$. The right plot (b) presents the residual values from \texttt{stella} as scatter points (with errorbars indicating the variation of the potential in the last 20\% of the time trace), the triangle marker shows the simulation of the flat-$B$ scenario, the solid line the numerical evaluation of Eq.~(\ref{eqn:general_RH_expression}), the dotted black line the analytical estimate of Xiao-Catto \citep{xiao2006short}, and the red dotted line the asymptotic expression in Eq.~(\ref{eqn:anal_lay_res}). The central bottom plot (c) shows the relative contribution to the residual by trapped/passing particles. The plots left and right represent the relative contribution to the residual by different parts of the population, where the vertical coordinate represents $1/\lambda$, with the black line representing $B$. The calculations are done at $k_\perp\rho_i\approx 0.048$ ($k_y\rho_i=0.05$ in \texttt{stella}).}
    \label{fig:sim_numerical}
\end{figure}
The behaviour of the residual at small mirror ratio can be checked against both careful numerical integration of Eq.~(\ref{eqn:general_RH_expression}) and linear electrostatic gyrokinetic simulations with the \texttt{stella} code \citep{barnes2019stella}. We present such a comparison in Figure~\ref{fig:sim_numerical}. For that comparison, a local field along a flux tube is constructed from a reference cyclone-base case (a simple Miller geometry \citep{miller1998noncircular}) whose $B$ has been modified with varying mirror ratios $\Delta$, while keeping all other elements of the geometry unchanged. The numerical evaluation of Eq.~(\ref{eqn:general_RH_expression}) is done by careful treatment of bounce integrals using double-exponential integration methods \citep{takahasi1974double} to appropriately deal with bounce points and logarithmic divergences in $\lambda$-space (details on the \texttt{python} code may be found in the Zenodo repository associated to this paper). The linear gyrokinetic simulations are run with large velocity space resolution in an attempt to resolve the boundary layer in velocity space to the best capacity within reason. This means that they must also be run for long times, on the order of the transit time of the smallest resolved velocity in order to reach the \textit{residual}. We take the residual from these simulations to be the value of the potential at the latest time simulated.\footnote{We are running these simulations in \texttt{stella} with $N_{v_\parallel}=2000$, $N_\mu=100$, $\Delta t=0.0125$ and $N_t=64000$, considered high resolutions. The smallest mirror ratio cases can be challenging to simulate and converge fully even under these extremely resolved conditions. For the semi-quantitative considerations in this paper we consider them to be sufficient, though. In addition to these numerical niceties, the physical oscillations of the electrostatic potential also pose an additional limitation, as these variations are not damped completely in the time domain of consideration for the lowest mirror ratios. This can lead to an inaccurate `measured' residual, but is once again deemed sufficient in the time domain considered for the semi-quantitative comparison here considered (see error-bars in Figure~\ref{fig:sim_numerical}).} Having these two numerical forms of assessing the residual provides us with additional forms to diagnose the results. In particular, and given the good agreement between the simulations with the numerical evaluation of the residual in Eq.~(\ref{eqn:general_RH_expression}), we can assess the contribution from different regions of velocity space to the residual using the latter (see Figure~\ref{fig:sim_numerical}c). 
\par
In the small mirror ratio limit, as predicted, there is a dominant contribution from a narrow boundary layer (group II). The analytic estimate of the residual in the small mirror ratio, Eq.~(\ref{eqn:anal_lay_res}), agrees to a good degree (within $\sim5-10\%$) with the simulation and integration (see red line in Figure~\ref{fig:sim_numerical}b). As the mirror ratio increases the importance within velocity space shifts (see Figure~\ref{fig:sim_numerical}c) and the bulk of trapped particles becomes dominant (the standard \cite{rosenbluth1998poloidal} picture). In that limit the residual can be estimated by \cite{rosenbluth1998poloidal} (RH),
\begin{equation}
    \left.\frac{\phi(\infty)}{\phi(0)}\right|_\mathrm{RH}=\frac{1}{1+1.6 \epsilon^2/(k_\perp\rho_i)^2\sqrt{\Delta}}, \label{eqn:residual_RH_typical}
\end{equation}
or more precisely by \cite{xiao2006short}, as explicitly shown in Fig.~\ref{fig:sim_numerical}b (black dotted line). The standard RH residual, Eq.~(\ref{eqn:residual_RH_typical}), exhibits a stronger dependence on the drift and transit time compared to the small mirror ratio limit, although the physical mechanism behind the residual remains broadly speaking the same. Namely, making the drift $\omega_d$ or the connection length smaller, the orbit width becomes smaller, so does the finite orbit polarisation and shielding power of the plasma, and thus the resulting residual grows.
\par
The preeminence of the RH or small mirror residual will change depending on the parameters of both the field and perturbation. A clear example of the latter is the dependence on $k_\perp\rho_i$. In fact, for any finite $\Delta$, there always exists a perpendicular length-scale long enough for which the RH scenario is recovered (formally, a value of $k_\psi$ below which the ordering $\epsilon^2\gg\Delta$ is violated), leading to a finite residual at small $k_\perp\rho_i$. Of course, the field parameters also play a key role. Most clearly, the variation of the mirror ratio $\Delta$ explicitly involves a regime transition between the $\Delta$-independent small-mirror residual, Eq.~(\ref{eqn:anal_lay_res}), and the RH residual (see Figure~\ref{fig:sim_numerical}b). This takes place when $\Delta\sim\epsilon^2$, which is approximately
\begin{equation}
    \Delta_t\approx 0.1(k_\perp\rho_i)^2\left(\frac{\omega_d/\omega_t}{k_\perp\rho_i}\right)^2. \label{eqn:Delta_transition_estimate}
\end{equation}
If the orbit width of the bulk is made larger, then the small-mirror contribution becomes relevant sooner. However, we must remain within the limit $\epsilon^2\ll1$, which we considered in the construction of our residual calculation. Staying within that limit, the transition mirror ratio must obey $\Delta_t<10^{-1}$, which implies that the transition occurs at small mirror ratios of at most a few per-cent. Of course, the exact value of this transition will generally not be as simple. We may compute it more accurately by defining numerically $\Delta_t$ as the mirror ratio at which the low $k_\perp\rho_i$ limit of the XC \citep{xiao2006short} residual matches the low-mirror ratio residual.
\par
Before moving to an analysis of these effects on different equilibria, let us turn to interpreting the time dependence of the residual observed in Figure~\ref{fig:sim_numerical}a. There are clearly two oscillation time-scales in the problem set-up considered: the faster damped geodesic-acoustic modes (GAMs) \citep{sugama2006collisionless,gao2006multiple,gao2008eigenmode,conway2021geodesic} and a slower oscillation. The former appear rather invariant under $\Delta$ (as one would expect from a passing ion dominated phenomenon), while the latter change significantly. 
In fact, this slower time scale behaviour is reminiscent of the slower oscillations attributed to the non-omnigeneous nature of stellarator fields \citep{mishchenko2008collisionless,mishchenko2012zonal,helander2011oscillations,monreal2017semianalytical,alonso2017observation}. This provides us with an additional way of interpreting the boundary layer contribution to the low-mirror residual. Because of their long transit time compared to their radial drift, these particles behave \textit{de facto} as non-omnigeneous particles, at least in a transient sense. The result are long time scale oscillations with a slow damping rate. The damping and frequency of oscillations grow in their time scale as $\Delta$ becomes smaller, which we attribute to the increasingly non-omnigeneous behaviour of the particles in this limit. A more in-depth investigation of this behaviour is left for future work.

\subsection{Geodesic acoustic mode (GAM) connection}
From the analysis of the time trace of our simulations, we observe that the residual and GAMs are just different dynamical phases of the same system. One then expects to see them both arise consistently in the same asymptotic limit. 
\par
GAMs are damped, oscillatory modes resulting from a balance between streaming and off-surface drift, basic reigning elements in the residual as well. Thus, these oscillatory modes are, like the residual, often studied as part of the assessment of the field response to zonal flows. The basic theoretical set-up for studying GAMs involves a flat-$|\mathbf{B}|$ field, where dynamics are dominated by passing ions, and the only inhomogeneity along field-lines is introduced by an oscillatory $\omega_d$. Under the assumption of a small $\omega_d/\omega_t$ (equivalent to the small $\epsilon$ we have considered in this paper), the behaviour of GAMs may be reduced to a simple dispersion relation \cite{sugama2005dynamics,sugama2006collisionless,gao2006multiple,gao2008eigenmode}. We reproduce some of the details of this derivation and the dispersion relation in Appendix~\ref{app:gam_derivation}. 
\par
The key observation is that the limit $\omega\rightarrow0$ of these dispersion relations, which determine the long time behaviour of the electrostatic potential \citep[Theorem.~2.36]{schiff2013laplace}, yields no residual. But we have shown just above that actually a finite residual remains in the limit of vanishing mirror ratio. A natural question thus arises: where is this residual hiding? It might be tempting to identify the slow GAM mode identified by \cite{gao2006multiple} with the residual, due to its similar form. This purely damped mode reads
\begin{equation}
        \frac{\phi(t\rightarrow\infty)}{\phi(0)}\approx\frac{1}{1+\frac{\epsilon^2}{4b}\left(1+\frac{\pi}{2(1+\tau)}\right)}e^{-\gamma t},
    \end{equation}
where,
\begin{equation}
    \frac{\gamma}{\omega_t}=\frac{\pi^{3/2}}{2}\left[\frac{2b}{\epsilon^2}+\left(\frac{1}{2}+\frac{\pi}{4(1+\tau)}\right)\right]^{-1}.
\end{equation}
The amplitude of the mode exhibits a quadratic finite orbit width dependence much in the fashion of the RH residual. Although the damping of the mode can be slow (with a characteristic decay time $\sim \epsilon^2/b\omega_t$), and thus display an effective value of the residual (transiently), it does not formally correspond to a collisionless, \textit{undamped} residual. 
\par 
In addition, it has a quadratic scaling rather than the linear one derived above. To resolve this apparent inconsistency we must recognise the importance of barely passing particles. For this subset of the population the transit time is so long that the ordering $\omega_t\gg\omega_d$ is not accurate, and thus the derivation of the usual GAM dispersion relation needs reworking. We present the details of how to do this in Appendix~\ref{app:gam_derivation}. Doing so, one can recover a finite valued residual with the same scaling as derived above, albeit with a different numerical factor. This difference is due to the difference in the derivation, and gives a factor of 0.20 instead of a 0.26 in Eq.~(\ref{eqn:residual_0p26}). This reconciling of the residual and GAM calculations is a theoretical relief.

\section{Field survey} \label{sec:field_survey}
In the preceding analysis of the residual problem we learned that there are two different regimes in which the behaviour of the residual is quite different. One, the regime where the layer dynamics become dominating, which occurs at small mirror ratios ($\Delta_t<10^{-1}$). And the more typical RH residual one, occurring at moderate values of $\Delta$, in which the bulk of the trapped particle population dominates the response of the system. We now explore the question of which regime prevails under the conditions that arise in different classes of magnetic equilibria.
\par
Let us start with the simplest family of magnetic field configurations: the circular tokamak. That is, an axisymmetric magnetic field configuration, with circular cross-sections and thus a unique magnetic well, which is the closest scenario to our idealised model-field. In such a scenario, we may reduce the relevant field properties to a few parameters, namely the safety factor $q$, the mirror ratio $\Delta$ and the radial wavenumber, $k_\perp\rho_i$. In the context of the residual, one may think of the safety factor $q$ as determining the ratio of the radial drift (in a tokamak $\omega_d\sim1/R$) to the connection length ($\omega_t^{-1}\sim qR$), explicitly $q=\omega_d/(\pi k_\perp\rho_i\omega_t)$. With that, the relevant expressions for the residual read, following Eqs.~(\ref{eqn:residual_0p26}) and (\ref{eqn:residual_RH_typical}),
\begin{equation}
    \left.\frac{\phi(\infty)}{\phi(0)}\right|_\mathrm{lay}\approx\frac{1}{1+1.63q/(k_\perp\rho_i)}, \quad \left.\frac{\phi(\infty)}{\phi(0)}\right|_\mathrm{RH}\approx\frac{1}{1+1.6 q^2/\sqrt{\Delta}}. \label{eq:tok-residuals}
\end{equation}
The larger the $q$, the larger the connection length, the larger the orbit width $\delta$ and the the lower the residual. In terms of these tokamak parameters, we may also rewrite the condition for the regime transition in Eq.~(\ref{eqn:Delta_transition_estimate}): the layer contribution becomes relevant for $\Delta<\Delta_t\sim(qk_\perp\rho_i)^2$. For a typical value of $q\sim1$, and a wavenumber $k_\perp\rho_i\sim0.1$, this implies mirror ratios below a percent. This is a rather small mirror ratio, which will only be reached sufficiently close to the magnetic axis (where $B$ is nearly constant due to axisymmetry). For shorter wavelengths or larger safety factors (which also reduce the residual) $\Delta_t$ will be larger. Because this occurs at the expense of larger orbit width, taking this limit to its extreme will ultimately lead to $\epsilon \sim 1$, implying $\delta>1$ for all particles, corresponding to a completely different regime.\footnote{Large wavenumber behaviour was explored by \citep{xiao2006short, monreal2016residual}. Physically, as the orbit sizes become large, they become less effective at shielding the original potential perturbation, and the residual grows. Note however that this large-$k_\perp\rho_i$ behaviour is more sensitive to initial conditions \citep{monreal2016residual} and electron dynamics should be brought in for a consistent treatment.} 


\par
To extend the discussion beyond the rather simplified case of circularly shaped tokamaks, we need some form in which to estimate the input parameters to our residual calculation.  We will focus on so-called optimised stellarator configurations: namely, \textit{quasisymmetric} \citep{boozer1983,nuhren1988,rodriguez2020} and \textit{quasi-isodynamic} \citep{Cary1997,Helander_2009,Nührenberg_2010} ones. The former can be seen as the natural generalisation of the axisymmetric case, where the field has a direction of symmetry on $|\mathbf{B}|$ instead of the whole vector $\mathbf{B}$. The direction of symmetry can be toroidal (\textit{quasi-axisymmetry}) or helical (\textit{quasi-helical}).
This symmetry forces the magnetic wells along the field line to be all nearly identical (same $B$ and $\omega_d$ \citep{boozer1983transport}, but different $k_\perp\rho_i$). In \textit{quasi-isodynamic} fields, the contours of $|\mathbf{B}|$ are closed poloidally, and carefully shaped to grant \textit{omnigeneity} \citep{bernardin1986,Cary1997,hall1975,helander2014theory}. As a result, wells are differently shaped, but all share the feature of being omnigeneous; that is, the orbits described by $\delta$ are closed as in Figure~\ref{fig:particle_orbits}. The description will in that case have to involve an average over wells. 
\par
Our approach now will be to construct effective model parameters for all of these configuration types, that may be applied to the above familiar expressions for the tokamak case, {\it e.g} Eqn.~\ref{eq:tok-residuals}.  These parameters will be derived using the inverse-coordinate near-axis description of equilibria \citep{garrenboozer1991a,landreman2019,rodriguez2023constructing,plunk2019direct}, as detailed in Appendix~\ref{app:nae}, and summarised in Table~\ref{tab:nae_geo}. We have included the case of a shaped tokamak for comparison. Let us now discuss the interpretation of these results.

\begin{table}
        \centering
        \begin{tabular}{c|c|c|c|}
                         & Tokamak & QS & QI \\ \hline
        $q_\mathrm{eff}$ & $ \displaystyle \frac{1}{\iota}\frac{\eta R_\mathrm{ax}}{\hat{\mathcal{G}}}$ & $\displaystyle \frac{1}{\iota-N}\frac{\eta R_\mathrm{ax}}{\hat{\mathcal{G}}}$ & $ \displaystyle \frac{2}{\pi N_\mathrm{nfp}}\frac{\bar{d}R_\mathrm{ax}}{\hat{\mathcal{G}}}$ \\
        $\Delta$ & $r\eta$ & $r\eta$ & $\Delta$ \\
        
        \end{tabular}
        \caption{\textbf{Characteristic near-axis residual-related parameters in optimised stellarators.} The table presents the value of the residual-relevant parameters $q_\mathrm{eff}$ and $\Delta$ for tokamaks and different optimised stellarator types, obtained using the near-axis description of the fields (see Appendix~\ref{app:nae}). The parameters are: $R_\mathrm{ax}$ the effective major radius (the length of the magnetic axis divided by $2\pi$), $\iota$ the rotational transform, $N$ the symmetry of the QS field, $N_\mathrm{nfp}$ number of field periods, $\eta$ and $\bar{d}$ leading poloidal variation of $|\mathbf{B}|$ over flux surfaces (roughly proportional to the axis curvature) and $\hat{\mathcal{G}}$ geometric factor defined in Eq.~(\ref{eqn:G_hat_def}).}
        \label{tab:nae_geo}
\end{table}
    
\begin{figure}
    \centering
    \includegraphics[width=\textwidth]{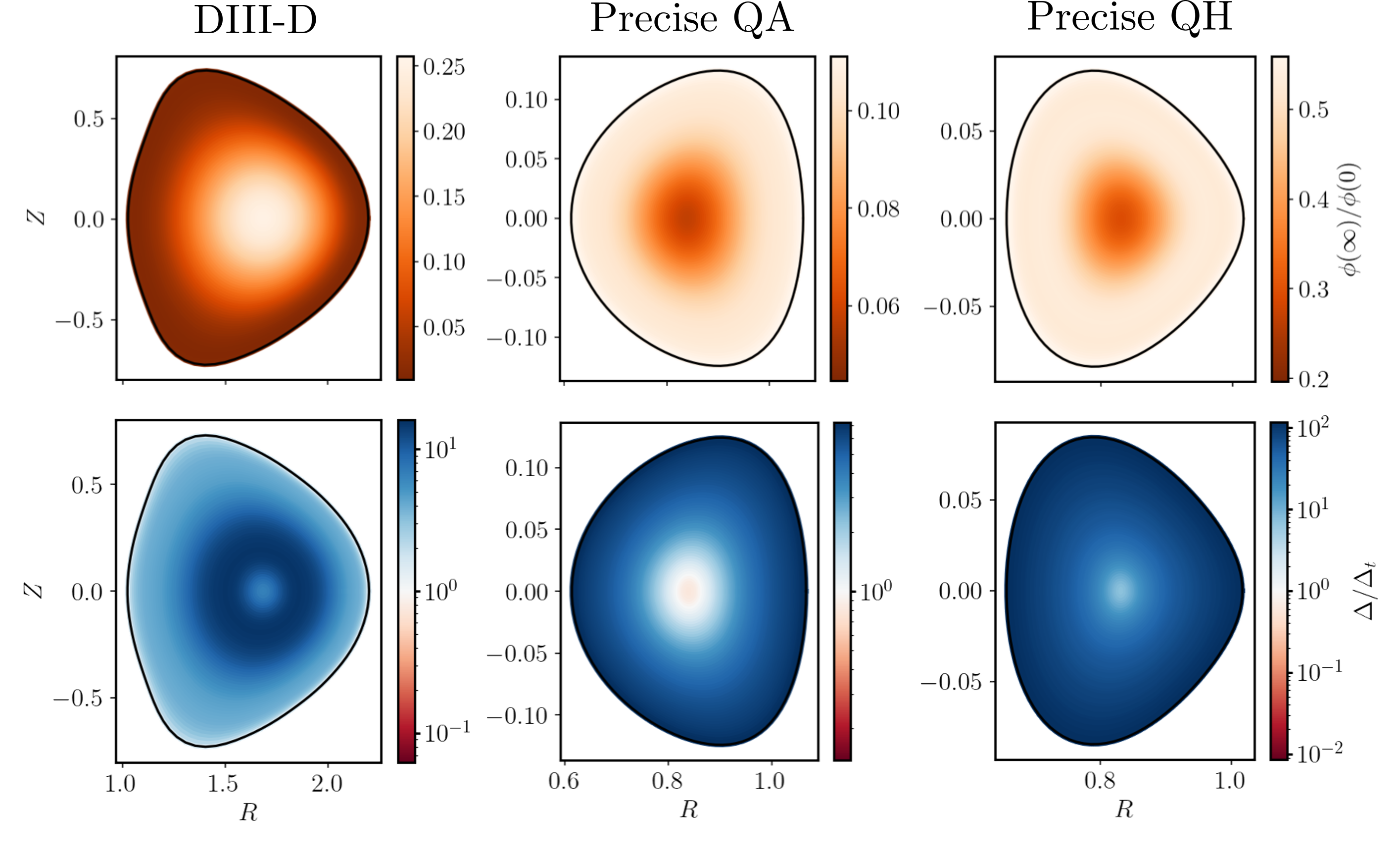}
    \caption{\textbf{Residual and closeness to the residual transition as a function of radius.} The plot shows the residual (top) and the ratio of the mirror ratio $\Delta$ to the residual regime tranition value $\Delta_t$ (bottom) for DIII-D (equilibrium from \cite{austin2019achievement}, shot 170680 at 2200ms) (tokamak), precise QA (QA stellarator) and precise QH (QH stellarator) configurations \citep{landreman2022magnetic}. The residual is computed numerically evaluating Eq.~(\ref{eqn:general_RH_expression}) using the global equilibria of the configurations to estimate the simplified single-well parameters for the residual calculation. The bottom plots are evaluated computing $\Delta_t$ as the mirror ratio value at which the XC estimate of the residual equals the small mirror ratio limit of the residual. It therefore is a measure of relevance of the low-mirror residual regime. It is clear that the centre of the QA configuration is where the low-mirror ratio is most relevant. The residual calculation was done for $k_\perp\rho_i=0.1$ for these.}
    \label{fig:radial_dep_residual}
\end{figure}

The first important distinction between fields is with regards to the behaviour of the mirror ratio. In tokamaks, as well as quasisymmetric stellarators, the mirror ratio has a strong radial dependence. In particular, because $|\mathbf{B}|$ has a direction of symmetry with a toroidal component, $\Delta$ must decrease towards the axis and do so at a rate related to the curvature of the field (within the near axis description it is proportional to the distance form the axis and $\eta\sim\kappa$, see Appendix~\ref{app:nae}). This implies the appearance of a finite region near the magnetic axis where the low-mirror residual becomes relevant. In practice, though, this region tends to be narrow, and thus likely unimportant (see Figure~\ref{fig:radial_dep_residual}s). 
It is particularly narrow in tokamaks, where the safety factor decreases towards the axis and can have a significant global shear, unlike quasisymmetric stellarators \citep{landreman2022magnetic,landreman2022mapping,rodriguez2023constructing,giuliani2024direct}. The consequence of this is also an inversion of the behaviour of the residual with radius: it tends to be largest in the core in a tokamak, but smallest for QS ones (see Figure~\ref{fig:radial_dep_residual}). QI stellarators are significantly different to both tokamaks and QS stellarators. As a result of having poloidally closed contours, the on-axis $|\mathbf{B}|$ is not constant, and thus the mirror ratio tends to a non-zero constant on the axis. This frees $\Delta$ from its strong radial dependence, preventing the low-mirror residual region from manifesting. 
\par
In addition to the differences in $\Delta$, the changes in the magnitude of the magnetic field gradient $\bnabla B$ (which affects $\omega_d$), the flux surface shaping (which affects $k_\perp$) and the connection length (which affects $\omega_t$) do also impact the residual. All of these physical elements may be captured in a parameter $q_\mathrm{eff}=\omega_d/(\pi k_\perp\rho_i\omega_t)$, given in Table~\ref{tab:nae_geo}. We define such a parameter to play the role that the safety factor takes in the circular-cross-section scenario of the residual. In particular, one should interpret this $q_\mathrm{eff}$ as a generalised form of $q$ in the residual expressed in Eq.~(\ref{eq:tok-residuals}) and other places. As such, larger $q_\mathrm{eff}$ implies lower residual and a higher relevance of the low-mirror residual regime.  Let us discuss what determines $q_\mathrm{eff}$ for each case in Table \ref{tab:nae_geo}.
\par
We start by analysing the role played by the perpendicular geometry (in particular $\langle|\nabla\psi|^2\rangle$). This is captured by (see Eqns.~\ref{eq:Ghat} and \ref{eq:Ghat-QI}),
\begin{equation}
    \hat{\mathcal{G}}^2=\frac{1}{2\pi}\int_0^{2\pi} \frac{\mathrm{d}\varphi}{\sin 2e}, \label{eqn:G_hat_def}
\end{equation}
where we define the angle $e$ such that $\mathcal{E}=\tan e$ is the elongation of the flux surfaces in the plane normal to the axis as a function of $\varphi$ \citep{rodriguez2023mhd} and we have considered the limit of small mirror ratio ($\Delta\ll1$). The angle $e\in(0,\pi/2)$ may be interpreted as the angle subtended by a right-angle triangle with the major and minor axes as catheti. Thus, a circular cross-section is represented by $e=\pi/4$, and the corresponding $\hat{\mathcal{G}}=1$. Any elliptical shape will then have a larger $\hat{\mathcal{G}}>1$ (as $\sin 2e<1$ for $e\neq \pi/4$ in the domain considered). Increasing the elongation of flux surfaces increases the average flux expansion, $\langle|\nabla\psi|^2\rangle$, leading to a decrease of $q_\mathrm{eff}$, a larger residual and a decrease in the importance of the low-mirror residual. This is consistent with \cite{xiao2007effects}. Physically, increasing elongation brings flux surfaces closer together, and thus narrows the orbit widths in real space. Any non-axisymmetric shape will necessarily have $\hat{\mathcal{G}}>1$ \citep{landreman2019,camacho-mata-2022,rodriguez2023mhd}, but variations between optimised configurations will be moderate given that limiting flux surface shaping is often an optimisation criterion. 
\par
Let us now focus on the differences in the magnitude of the magnetic drifts. The drift is controlled by the gradients of $|\mathbf{B}|$, which decrease the residual the larger they become. The balance between magnetic gradients (and thus magnetic pressure) and magnetic field line tension provides an important observation: the more curved field lines are, the stronger the gradients. In the near axis framework, this naturally leads to a picture in which the more strongly shaped a magnetic axis is, the larger the gradients will be. This behaviour is represented by parameters $\eta$ and $\bar{d}$ in Table~\ref{tab:nae_geo} (see Appendix~\ref{app:nae} for a more precise description), which typically scale like $\eta\sim\kappa$ \citep{rodriguez2023constructing}, where $\kappa$ is the axis curvature. For similarly shaped cross-sections, $\eta$ (or $\bar{d}$) will be larger for QH and QI stellarators compared to QA and tokamaks \citep{rodriguez2022phases, camacho-mata-2022}, and even more with the number of field periods. The drift in the QI case deserves special consideration, because the pointwise radial drift varies from field line to field line, vanishing on some \citep{Helander_2009,landreman2012}. Thus, on `average', the drift in these configurations is smaller (see Appendix~\ref{app:nae} for the details), which can enhance the residual. In brief, QH configurations are expected to have the largest field gradients, followed by QIs in which the field-line averaging reduces the effective gradients, and finally QAs and tokamaks. 
\par
The last element of consideration in $q_\mathrm{eff}$ is the connection length, i.e. the length along the field line of a magnetic well. The difference in the topology of the $|\mathbf{B}|$ contours (and their alignment to magnetic field lines) leads to the following comparative scaling, $R_\mathrm{ax}/\iota : R_\mathrm{ax}/(\iota-N) : R_\mathrm{ax}/N_\mathrm{nfp}$. Of course, this naturally leads to ordering the connection lengths to be largest for QA and tokamaks, smaller for QHs and the smallest for QIs. This follows from the observation that the number of field periods serves as an upper bound of $\iota$ for QHs in practice. 
\par
The three elements discussed above compete with each other, but the preeminence of the connection length on $q_\mathrm{eff}$ in practice leads to the relative ordering,
\begin{equation}
    q_{\mathrm{eff, tok}}\sim q_{\mathrm{eff, QA}}>q_{\mathrm{eff, QH}}\gtrsim q_{\mathrm{eff, QI}}.
\end{equation}
This should be regarded as a rough guide, not as a rigid rule; a similar ordering for the overall size of the residual is argued by \cite{plunk2024residual}. 
\begin{figure}
    \centering
    \includegraphics[width=\textwidth]{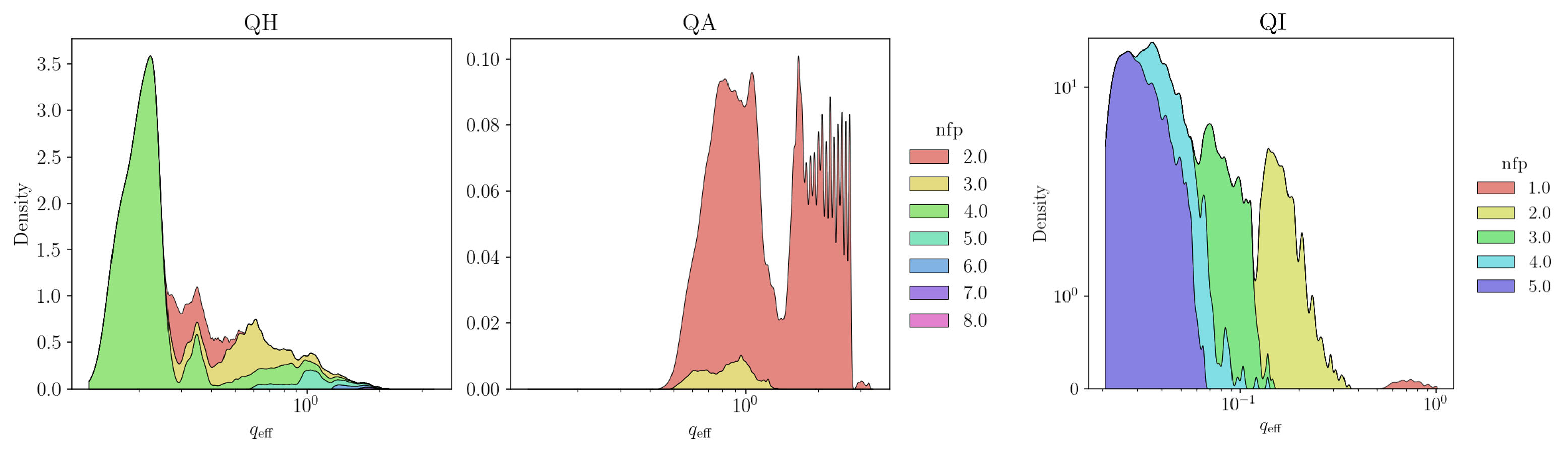}
    \caption{\textbf{Parameter $q_\mathrm{eff}$ for QS and QI configurations.} Statistics of $q_\mathrm{eff}$ for QS and QI configurations. The left plots represent the normalised (by total area) density of QH and QA configurations by their value of $q_\mathrm{eff}$ in the QS near-axis database in \cite{landreman2022mapping}, which serves as a representative population of optimised QS configurations. The density for each number of field period (color) is stacked vertically on top of one another, and represents the number of configurations in the database satisfying those parameters. The rightmost plot shows the same analysis through a QI near-axis database \citep{plunk2024-QI}. This shows the rough relative ordering of $q_\mathrm{eff}$ between different omnigeneous fields, as indicated in the text. Most QH configurations are $N=4$, and their $q_\mathrm{eff}$ is the lowest for all $N$, while larger or smaller $N$ lead roughly to larger $q_\mathrm{eff}$. This shows the complexity and detail of  The $N=2$ is the main QA. }
    \label{fig:enter-label}
\end{figure}
To strengthen and illustrate this behaviour of $q_\mathrm{eff}$ across different configurations, we use the large database of near-axis QS configurations of \cite{landreman2022mapping} and near-axis QI configurations of \cite{plunk2024-QI} to evaluate this parameter across configurations. This confirms that one expects the residual to be smallest in tokamaks and QAs, with the small-mirror regime barely becoming relevant near their core. We leave a more complete analysis of these databases and the lessons to be learned from these for the future. We also note that more complex field shaping beyond the simple model used in this paper could change some of the exact quantitative behaviour observed concerning especially the location of the residual transition, but we also leave this to future investigations.
\par

\section{Conclusions}
In this paper, we have carefully analysed the behaviour of the residual in the limit of small mirror ratio. The contribution of barely passing particles provides a finite residual in this limit, changing its usual scaling and exchanging roles of the importance between trapped and passing particles. We identify the role of such barely trapped particles and provide some analytical estimates, that we compare to some gyrokinetic simulations. This limiting behaviour, however, is shown to occur at very small mirror ratios $\Delta<(\omega_d/\omega_t)^2$, where $\omega_d$ is the radial drift frequency and $\omega_t$ the transit frequency of a thermal particle to travel a connection length. An analysis using near-axis theory of this effect through tokamaks, quasisymmetric and quasi-isodynamic stellarators suggests that although barely, the centre of quasi-axisymmetric stellarators is the region in which some of these effects could manifest most clearly. This analysis also shows (including a cross-check through a large database of configurations) that the residual itself tends to be larger in quasi-isodynamic stellarators, to be followed by quasi-helical and lastly quasi-axisymmetric (and tokamak) ones.

\section*{Data availability}
The data that support the findings of this study are openly available at the Zenodo repository with DOI/URL 10.5281/zenodo.12805697.

\section*{Acknowledgements}
We gratefully acknowledge fruitful discussion with R. Nies and W. Sengupta. 

\section*{Funding}
E. R. was supported by a grant of the Alexander-von-Humboldt-Stiftung, Bonn, Germany, through a postdoctoral research fellowship. 

\section*{Declaration of interest}
The authors report no conflict of interest.

\appendix

\section{Additional details on the orbit widths} \label{app:orbit_width}
In this appendix we complete the information about the finite orbit width provided in Section~\ref{sec:orbit_width}, necessary to complete the residual calculation in Section~\ref{sec:derivation_small_mirror_residual}.

\subsection{Passing particles}
Let us consider the shape of the orbits described by the barely passing particles living within the boundary layer defined in Section~\ref{sec:passing_particles} (see Figure~\ref{fig:particle_orbits}). To evaluate the residual integrals in Eq.~(\ref{eqn:general_RH_expression}) we require information about the turning points of $\delta$. In particular, besides the location and value of $\delta$ extrema, the second derivative \citep[Sec.~6.5]{bender2013advanced}. The second derivative at those points is,  
\begin{equation}
    \delta''_\mathrm{pass}=\sigma \frac{v}{v_T} \frac{\epsilon\pi^2}{2}\times
    \begin{cases}
    \begin{aligned}
        & \frac{1}{\sqrt{\hat{\lambda}}}, & \quad (\bar{\ell}=\pm1) \\
        & -\frac{1}{\sqrt{2\Delta+\hat{\lambda}}}, & \quad (\bar{\ell}=0)
    \end{aligned}
    \end{cases}
\end{equation}
where we have used the definition of $\hat{\lambda}$ and $\Delta\ll1$.
\par
To complete the orbit description, we also need the transit time of passing particles. In the simplified single well model, this is defined to be the time taken by a particle to move from $\bar{\ell}=-1$ to $1$. The time can be expressed \citep[Eq.~(7.27)]{helander2005collisional} in terms of the elliptic function $K$ \citep[Sec.~19]{DLMF}\citep[Eq.~(16.1.1)]{abramowitz1968handbook},
\begin{equation}
    \tau_t \omega_t=\frac{4}{\pi}\frac{v_T}{v}\frac{K(\kappa)}{\sqrt{1-\lambda\bar{B}(1-\Delta)}},
\end{equation}
where $\kappa=2\lambda\Delta/[1/\bar{B}-\lambda(1-\Delta)]$.

\subsection{Trapped particles}
The orbits described by trapped particles are ostensibly different. The function $\delta(\bar{\ell})$ has a single turning point at the centre of the orbit, point at which the second derivative is
\begin{equation}
    \delta_\mathrm{trap}''(0)\approx\sigma \frac{v}{v_T}\frac{\epsilon\pi^2}{\sqrt{2\bar{\kappa}\Delta}}.
\end{equation}
The orbits, unlike those of passing particles, are sharp at, in this case, bounce points. This is a result of the particles spending longer at these points, where the radial drift is non-zero. This difference in how particles spend their time on different parts of their orbit also affects the expression for the orbit time, here called bounce time \citep{connor1983effect}\citep[Eq.~(7.28)]{helander2005collisional}, 
\begin{equation}
    \tau_b\omega_t=\frac{2}{\pi}\frac{v_T}{v}\sqrt{\frac{2}{\lambda\bar{B}\Delta}}K(\bar{\kappa}).
\end{equation}

\section{Residual in a GAM scenario} \label{app:gam_derivation}
In this Appendix we present how the description of geodesic acoustic modes (GAMs) can be made to align with the finite residual result derived in the main text. To that end, let us start by re-writing the linearised gyrokinetic equation in Eq.~(\ref{eqn:GK}) and dropping the initial condition,
\begin{equation}
    iv_\parallel\partial_\ell \hat{g}+(\omega-\tilde{\omega}_d)\hat{g}-J_0F_0\omega \frac{q\hat{\phi}}{T}=0. \label{eqn:gk}
\end{equation}
As in the residual calculation, we have written the equation for $k_\alpha=0$, which leads to vanishing of the diamagnetic drive.
\par
Because we are here interested in the GAM dynamics, it is conventional to specialise to an artificial flat-$B$ field, one in which the sole field property that varies along the field-line is the curvature drift (i.e. $k_\perp\rho_i$ is also constant). Modelling $\omega_d(\ell)=\omega_d\cos(\pi\ell/L_d)$, we may Fourier resolve Eq.~(\ref{eqn:gk}) writing $\hat{g}=\sum_{n=-\infty}^\infty \hat{g}_ne^{in\pi\ell/L_d}$ and $\hat{\phi}=\sum_{n=-\infty}^\infty \hat{\phi}_ne^{in \pi \ell/L_d}$. Taking into account the coupling through $\omega_d$, and
\begin{equation}
    \hat{g} \cos\left(\frac{\pi\ell}{L_d}\right)=\frac{1}{2}\sum_{n=-\infty}^\infty (\hat{g}_{n+1}+\hat{g}_{n-1})e^{in\pi\ell/L_d}, 
\end{equation}
we may then write Eq.~(\ref{eqn:gk}) as,
\begin{equation}
    \left(-nx_\parallel+\frac{\omega}{\omega_t}\right) \hat{g}_n-\frac{\tilde{\omega}_d}{2\omega_t}\left(\hat{g}_{n-1}+\hat{g}_{n+1}\right)=F_0J_0\frac{\omega}{\omega_t}\frac{q\hat{\phi}_n}{T},
\end{equation}
where $\omega_t=\pi v_T/L_d$ is the transit frequency over the characteristic scale of the drift variation and $x_\parallel=v_\parallel/v_T$. 
\par
The system has a sideband coupling through the drift, whose overlap is controlled by $\omega_d/\omega_t$. Thus, ordering $\epsilon=\omega_d/\omega_t\ll1$ is particularly convenient to regularise the problem and be able to truncate it. In fact, if we drive the system uniformly, meaning we assume $\hat{\phi}_0,~\hat{g}_0\sim O(1)$, we expect to find small sidebands. That way, we may focus on the following reduced system of equations,
\begin{subequations}
\begin{align}
    \left(x_\parallel+\frac{\omega}{\omega_t}\right)\hat{g}_{-1}-\frac{\tilde{\omega}_d}{2\omega_t}\hat{g}_0\approx F_0J_0\frac{\omega}{\omega_t}\frac{q\hat{\phi}_{-1}}{T}, \\
    \frac{\omega}{\omega_t}\hat{g}_{0}-\frac{\tilde{\omega}_d}{2\omega_t}(\hat{g}_{-1}+\hat{g}_{1})\approx F_0J_0\frac{\omega}{\omega_t}\frac{q\hat{\phi}_{0}}{T},\label{eqn:g_0_eqn} \\
    -\left(x_\parallel-\frac{\omega}{\omega_t}\right)\hat{g}_{1}-\frac{\tilde{\omega}_d}{2\omega_t}\hat{g}_0\approx F_0J_0\frac{\omega}{\omega_t}\frac{q\hat{\phi}_{1}}{T}.
\end{align} \label{eqn:system_eqn_GK_fourier}
\end{subequations}
In addition to the gyrokinetic equation written in this form, we must complete the eigenvalue problem with the quasineutrality condition. The condition, now explicitly involving electrons ($e$) and ions ($i$), reads in this basis,
\begin{equation}
    \frac{T_i}{q_i}\sum_{s=e,i}\int J_{0s} \hat{g}_{s,k} \mathrm{d}^3\mathbf{v}=n(1+\tau)\hat{\phi}_k,
\end{equation}
where the sum is over both ions and electrons.  To construct the final form of the dispersion we shall eventually use $b_e/b_i\sim m_e/m_i\ll 1$, $\zeta_e/\zeta_i\sim\sqrt{m_e/m_i}\ll1$ and $\epsilon_e/\epsilon_i\sim\sqrt{m_i/m_e}$.

\subsection{GAM dispersion}
The common form of the dispersion relation for GAMs is obtained by combining the equations in Eqs.~(\ref{eqn:system_eqn_GK_fourier}) to write $\hat{g}_0$ explicitly as function of $\hat{\phi}_0$ to leading order in $O(\epsilon^2)$ and performing the appropriate velocity space integrals. The result \citep{gao2006multiple,gao2008eigenmode,sugama2006collisionless},
\begin{equation}
    \mathcal{D}=1-\Gamma_0(b)+\frac{\epsilon^2}{2}\left[\mathcal{D}^{(2)}-\frac{(\mathcal{D}^{(1)})^2}{1+\tau+\mathcal{D}^{(0)}}\right], \label{eqn:gam_disp_four}
\end{equation}
where,
\begin{subequations}    
\begin{gather}
        \mathcal{D}^{(2)}= ~\frac{1}{\zeta}\left[\Gamma_0(b)\frac{\zeta}{2}\left(1+2\zeta^2(1+\zeta Z(\zeta))\right)+F_2(b)\zeta(1+\zeta Z(\zeta))+\frac{1}{4}F_4(b)Z(\zeta)\right], \\
        \mathcal{D}^{(1)}= ~\Gamma_0(b)\zeta(1+\zeta Z(\zeta))+\frac{1}{2}F_2(b)Z(\zeta), \\
        \mathcal{D}^{(0)}= \Gamma_0(b)\zeta Z(\zeta),
    \end{gather}
    \begin{equation}
        \int F_0J_0^2\mathrm{d}^3\mathbf{v}= \Gamma_0(b), \label{eqn:int_simp}
    \end{equation}
\end{subequations}
and $\zeta=\omega/\omega_t$. The dispersion relation is consistent with multiple modes, which have been explored in \cite{gao2008eigenmode}. Note that in those pieces of work \citep{gao2006multiple,gao2008eigenmode,sugama2006collisionless}, the problem is solved not using a Fourier resolution of the problem like we have here, but instead using the integrating factor approach of \cite{connor1980stability}. 
\par
The dispersion relation in Eq.~(\ref{eqn:gam_disp_four}) can be assessed near $\zeta\rightarrow 0$, which is responsible for the long time response of the plasma \citep[Theorem~2.36]{schiff2013laplace}. It may be shown by expanding the dispersion function \citep{fried2015plasma}, and taking for simplicity the small finite Larmor radius limit, 
\begin{equation}
    \mathcal{D}\approx \frac{b}{\omega}\left[1+\frac{\epsilon^2}{4b}\left(1+\frac{\pi}{2(1+\tau)}\right)\right]\left(\omega-\omega_0\right),
\end{equation}
where
\begin{equation}
    \frac{\omega_0}{\omega_t}=-i\frac{\sqrt{\pi}}{2}\left[\frac{2b}{\epsilon^2}+\left(\frac{1}{2}+\frac{\pi}{4(1+\tau)}\right)\right]^{-1}.
\end{equation}
The system shows a purely damped mode, but no truly net residual. 

\subsection{Revival of the residual}
This \textit{no residual} conclusion is not consistent with the calculation in this paper. So, where is the residual hiding? To see how the approach to the GAM could have missed the residual contribution, let us go back to the truncated system of equations where the $n=0,~\pm1$ modes are retained, Eqs.~(\ref{eqn:system_eqn_GK_fourier}), and recombine them into
\begin{subequations}
    \begin{multline}
        \frac{T/q}{F_0J_0}g_\pm=\frac{1}{2}\frac{(\tilde{\omega}_d/\omega_t)^2\mp4\zeta(x_\parallel\pm\zeta)}{(\tilde{\omega}_d/\omega_t)^2+2(x_\parallel^2-\zeta^2)}\phi_\pm \mp \frac{\tilde{\omega}_d}{\omega_t}\frac{x_\parallel\pm\zeta}{(\tilde{\omega}_d/\omega_t)^2+2(x_\parallel^2-\zeta^2)}\phi_0\\
        -\frac{1}{2}\left(\frac{\tilde{\omega}_d}{\omega_t}\right)^2\frac{1}{(\tilde{\omega}_d/\omega_t)^2+2(x_\parallel^2-\zeta^2)}\phi_\mp, \label{eqn:gen_four_g_pm}
    \end{multline}
    \begin{multline}
        \frac{T/q}{F_0J_0}g_0=\frac{2(x_\parallel^2-\zeta^2)}{(\tilde{\omega}_d/\omega_t)^2+2(x_\parallel^2-\zeta^2)}\phi_0+\frac{\tilde{\omega}_d}{\omega_t}\frac{x_\parallel-\zeta}{(\tilde{\omega}_d/\omega_t)^2+2(x_\parallel^2-\zeta^2)}\phi_-\\
        -\frac{\tilde{\omega}_d}{\omega_t}\frac{x_\parallel+\zeta}{(\tilde{\omega}_d/\omega_t)^2+2(x_\parallel^2-\zeta^2)}\phi_+,\label{eqn:gen_four_g_0}
    \end{multline} \label{eqn:gk_four_trunc}
\end{subequations}
where $\pm$ denote the $n=\pm1$ sidebands. We did not use this full form of the equations when deriving the dispersion relation for the GAMs, but instead their limit when $\epsilon=\omega_d/\omega\ll1$. Formally, this ordering was used to expand the kinetic resonant denominators
\begin{equation}
    \mathcal{R}=\frac{1}{\tilde{\omega}_d^2/\omega_t^2+2(x_\parallel^2-\zeta^2)},
\end{equation}
that are found ubiquitous in Eqs.~(\ref{eqn:gk_four_trunc}). For this expansion in the denominator to be sound we must have, of course, $x_\parallel^2-\zeta^2\gg\tilde{\omega}_d^2/\omega_t^2$, where we shall not forget the velocity space dependence of $\tilde{\omega}_d=\omega_d(x_\parallel^2+x_\perp^2/2)$. The GAM dispersion relation thus fails to describe any physics where $x_\parallel^2-\zeta^2\ll \epsilon^2 x_\perp^4/4$. This is especially problematic at long time scales (i.e. within a layer in $\omega$-space where $\omega<\omega_d$) and for the part of the population living within a narrow layer of order $x_\parallel\sim\epsilon$ in velocity space near $x_\parallel=0$. I.e. the GAM description overlooks the contribution from \textit{barely passing} particles, whose transit time is significantly longer than that of the bulk. 
\par
The question is then, how can one capture the behaviour from within this layer properly in this GAM formalism? Can one recover a residual result like that in Eq.~(\ref{eqn:residual_0p26})? To do so we must not expand in small $\tilde{\omega}_d$, but instead do so in $\zeta\rightarrow 0^+$ (indicating approach from the positive $\Im\{\omega\}$ direction). With this in mind, let us write the quasineutrality condition applied to Eq.~(\ref{eqn:gen_four_g_0}) as
\begin{equation}
    (1+\tau-\mathcal{D}^{(2)})\hat{\phi}(0)\approx -\frac{\epsilon}{2}\left[\mathcal{D}^{(1)}_-\hat{\phi}(-1)-\mathcal{D}^{(1)}_+\hat{\phi}(1)\right], \label{eqn:barely_gam_res_I}
\end{equation}
where
\begin{align}
    \mathcal{D}^{(2)}=&~ \frac{2}{\bar{n}}\int F_0J_0^2(x_\parallel^2-\zeta^2)\mathcal{R} \mathrm{d}^3\mathbf{v} \label{eqn:D2_gam_res}
\end{align}
\begin{align}
    \mathcal{D}^{(1)}_\pm= -\frac{2}{\bar{n}}\int F_0J_0^2\left(x_\parallel^2+\frac{x_\perp^2}{2}\right)(x_\parallel\pm\zeta)\mathcal{R} \mathrm{d}^3\mathbf{v}. \label{eqn:D1_gam_res}
\end{align}
To evaluate these integrals, we rewrite $\mathcal{R}$ by separating it into a sum over simple poles. To do so, we define, 
    \begin{equation}
        \Delta=~ \sqrt{\frac{1}{\epsilon^2}+x_\perp^2+2\zeta^2}, \quad \zeta_\pm=~ \frac{\Delta}{\epsilon}\pm\left(\frac{1}{\epsilon^2}+\frac{x_\perp^2}{2}\right),
    \end{equation}
so that 
\begin{equation}
    \mathcal{R}=-\frac{1}{2\epsilon\Delta}\left[\frac{1}{x_\parallel^2+\zeta_+}-\frac{1}{x_\parallel^2-\zeta_-}\right].
\end{equation}
Choosing the negative branch of the square root for a correct continuation from $\Im\{\zeta\}>0$ to the rest of the complex plane,
\begin{equation}
    \frac{1}{x_\parallel^2\pm\zeta_\pm}=\frac{1}{2\sqrt{\mp\zeta_\pm}}\left(\frac{1}{x_\parallel-\sqrt{\mp\zeta_\pm}}-\frac{1}{x_\parallel+\sqrt{\mp\zeta_\pm}}\right), \label{eqn:split_pre_Z}
\end{equation}
in such a way that the integrals Eqs.~(\ref{eqn:D2_gam_res})-(\ref{eqn:D1_gam_res}) explicitly involve integrals over $x_\parallel$. This form of $\mathcal{R}$ allows us to express integrals in terms of plasma dispersion functions \citep{fried2015plasma} upon appropriate redefinition of the sign of $x_\parallel$ (which will annihilate the contribution from odd $x_\parallel$ terms).\footnote{We shall here not be extremely careful with the definition of branch cuts and the precise deformation of the Laplace contour in $\zeta$-space. This would be needed for a fuller description of the time response of the system (one that captures the contribution from branch cuts for example), but here we content ourselves with the $\zeta\rightarrow0$ response.} As a result, we may write the integrals as a combination of 
\begin{align}
    I_{nm}=&~ \frac{1}{\bar{n}}\int x_\parallel^{2n}x_\perp^{2m}F_0J_0^2\mathcal{R}\mathrm{d}^3\mathbf{v}=-\frac{1}{\epsilon}\int_0^\infty x_\perp^{2m+1} J_0^2e^{-x_\perp^2}\frac{1}{\Delta}\left[\frac{Z_n(\sqrt{-\zeta_+})}{\sqrt{-\zeta_+}}-\frac{Z_n(\sqrt{\zeta_-})}{\sqrt{\zeta_-}}\right]\mathrm{d}x_\perp,
\end{align}
where we define,
\begin{equation}
    Z_n(x)=\frac{1}{\sqrt{\pi}}\int_{-\infty}^\infty\frac{x_\parallel^{2n}e^{-x_\parallel^2}}{x_\parallel-x}\mathrm{d}x_\parallel,
\end{equation}
for $\Im\{x\}>0$, and analytically continued to the rest of the complex plane. In particular, we may write
\begin{subequations}
    \begin{align}
        \mathcal{D}_\pm^{(1)}=\mp 2\zeta\left(I_{10}+\frac{I_{01}}{2}\right), \label{eqn:D1_gam_lay}\\
        \mathcal{D}^{(2)}=2\left(I_{10}-\zeta^2 I_{00}\right). \label{eqn:D2_gam_lay}
    \end{align}
\end{subequations}
\par
These integrals remain quite sophisticated, and simplifying them is paramount to analytically proceed forward. A natural simplifying attempt is to use asymptotic forms of the plasma dispersion function \citep{fried2015plasma}. The argument $\zeta_+\approx 2/\epsilon^2+x_\perp^2-\epsilon^2 x_\perp^4/8$, which is a large and positive real part quantity owing to the largeness of $1/\epsilon^2$, we may use the asymptotic form \citep[Sec.~IID]{fried2015plasma} $Z(x)\approx-\sum_{n=0}^\infty x^{-(2n+1)}(n-1/2)!/\sqrt{\pi}$ (the exponential term is exponentially small). In the case of $\zeta_-\approx \zeta^2-x_\perp^4\epsilon^2/8$ and we may consider an expansion in this small argument. Namely, \citep[Sec.~IIC]{fried2015plasma} $Z(x)=i\sqrt{\pi}\exp(-x^2)-x\sum_{n=0}^\infty (-x^2)^n\sqrt{\pi}/(n+1/2)!$. This introduces a leading order non-zero imaginary contribution. 
\par
With the above tools in place, we may proceed and compute the required integrals to the necessary order. 
\subsubsection{Integrals for $\mathcal{D}^{(2)}$}
Let us compute first the leading order $I_{00}$. Without having to go into the complex details about the specific branch cuts and complex quadrant of $\zeta$ in the complex plane, one can show \citep[Eq.~3.387.7]{gradshteyn2014table}
\begin{equation}
    I_{00}\approx~ \int_0^\infty x_\perp J_0^2 e^{-x_\perp^2}\frac{\sqrt{\pi}}{\sqrt{\frac{x_\perp^4\epsilon^2}{8}-\zeta^2}}\mathrm{d}x_\perp[1+O(\zeta,\epsilon^2)] \propto \frac{1}{\epsilon}\ln\left(\frac{\epsilon}{\zeta\sqrt{2}}\right),
\end{equation}
where for this estimate we have assumed $b\ll1$ to approximate $J_0\sim1$ and we have kept the leading order term in $\zeta$ (in the limit of small $\zeta$). So, in the limit of $\zeta\rightarrow0$, this integral diverges logarithmically, but its contribution to $\mathcal{D}^{(2)}$ vanishes, Eq.~(\ref{eqn:D2_gam_lay}). 
\par
Computing then $I_{10}$, and using $Z_1(x)=x[1+xZ(x)]$, 
\begin{align}
    I_{10}\approx&~ -\int_0^\infty x_\perp J_0^2 e^{-x_\perp^2}\left[-1+\frac{\epsilon}{2}\sqrt{\frac{\pi}{2}}x_\perp^2+\frac{\epsilon^2}{4}(1+2x_\perp^2-x_\perp^4)+O(\epsilon^3)\right] \mathrm{d}x_\perp \\
    \approx&~ \frac{1}{2}\left[\Gamma_0(b)-\frac{\epsilon}{2}\sqrt{\frac{\pi}{2}}F_2(b)-\frac{\epsilon^2}{4}(\Gamma_0(b)+2F_2(b)-F_4(b))\right]\\
    \approx&~ -\frac{1}{2}\left(b-1+\frac{\epsilon}{2}\sqrt{\frac{\pi}{2}}+\frac{\epsilon^2}{4}\right)=\frac{1}{2}\mathcal{D}^{(2)},
\end{align}
where we used the relevant Weber integrals \citep[Eq.~6.615]{gradshteyn2014table} and the notation $F_n=2\int_0^\infty x^{n+1}e^{-x^2}J_0^2(x\sqrt{2b})\mathrm{d}x$, and in the last line considered the small $b$ limit. Importantly, there is a term linear in $\epsilon$ which comes from the pole contribution to the plasma dispersion function. 
\par

\subsubsection{Integrals for $\mathcal{D}_\pm^{(1)}$}
With $\mathcal{D}^{(2)}$ constructed, we may turn to $\mathcal{D}^{(1)}$, Eq.~(\ref{eqn:D1_gam_lay}). The integral has an overall factor of $\zeta$, and thus to leading order, it will vanish unless there is some $\zeta$-divergence. The term $I_{10}$, which we have just computed, does not have such divergence, and thus its contribution will vanish. So we only need to calculate $I_{01}$, which one may show to be $I_{01}\approx\sqrt{2}/\epsilon$ to leading order. Thus, $\mathcal{D}^{(1)}_\pm\sim O(\zeta)$, and thus it will vanish in the small $\zeta$ limit. One may savely drop the coupling terms in Eq.~(\ref{eqn:barely_gam_res_I}) (the sideband $\phi_\pm$ does not have any divergent behaviour neither).
\par
\subsubsection{Dispersion relation}
Thus, the remaining dispersion function is,
\begin{equation}
    \mathcal{D}=1-\left[\Gamma_0(b)-\frac{\epsilon}{2}\sqrt{\frac{\pi}{2}}F_2(b)-\frac{\epsilon^2}{4}(\Gamma_0(b)+2F_2(b)-F_4(b))\right],
\end{equation}
where we have summed over species and taken the limit of $m_e/m_i\ll1$, and all quantities here should now be considered to represent ions. The value of the residual can then be written\footnote{We are being loose here about initial condition, but we may simply consider the RH initial condition of a uniformly perturbed potential.}, assuming $b\ll1$ for simplicity,
\begin{equation}
    \frac{\phi(\infty)}{\phi(0)}=\frac{1}{1+\frac{\epsilon}{2b}\sqrt{\frac{\pi}{2}}+\frac{\epsilon^2}{4b}}.
\end{equation}
It includes the leading order linear term in $\epsilon$, as the residual expression in the main text does. The difference with the result in the main text is the numerical factor in front of the linear term. As opposed to the $0.26 \omega_d/(b\omega_t)$ obtained in the text, and realising that $\omega_t$ as used in this appendix is $\pi$ times that in the main text, the result here yields $(1/2\sqrt{2\pi})\omega_d/(b\omega_t)\approx 0.20\omega_d/(b\omega_t)$. This is a 30\% discrepancy between both estimates of the residual, but the same scaling nonetheless.

\section{Near-axis properties in optimised configurations} \label{app:nae}
In this Appendix we present the near-axis calculations necessary to obtain the expressions in Table~\ref{tab:nae_geo} for the residual relevant parameters in different omnigeneous magnetic fields. These should be taken as informed estimates for the amplitudes of the simple model assumed in the main text. As we shall show, this is a good fit for QS fields, but not so much for QI. We assume some basic understanding of inverse-coordinate near-axis theory \citep{garrenboozer1991a, garrenboozer1991b}, and shall not derive the basic building elements of it. We refer the reader to the work by \cite{landreman2019} for the general equations for magnetohydrostatic equilibrium and in particular in a quasisymmetric configuration, and \cite{plunk2019direct,rodriguez2023higher} for quasi-isodynamic ones. We shall here use, with further explicit reference to those works, the elements needed for the evaluation of the appropriate quantities.

\subsection{Quasisymmetric fields}
Let us start by writing the magnetic field magnitude near the axis for a quasisymmetric field \citep[Eq.~(A1)]{garrenboozer1991b} \citep[Eq.~(2.15)]{landreman2019},
\begin{equation}
    B\approx B_0(1+r\eta\cos\chi), \label{eqn:nae_qs_B}
\end{equation}
where $r=\sqrt{2\psi/\bar{B}}$ is a pseudo-radial coordinate normalised to a reference $\bar{B}$, and $\chi=\theta-N\varphi$, where $N$ is the direction of symmetry of the QS field and we are using Boozer coordinates. Because $B_0$ is a constant, it is clear from this form that the constant parameter $\eta$ measures the variation of the magnetic field within a surface (to leading order). Thus, along a field line (at constant $\alpha$) the magnetic field depends on $\chi=\alpha+\bar{\iota}\varphi$, and thus the mirror ratio is,
\begin{equation}
    \Delta=r\eta,
\end{equation}
as indicated in Table~\ref{tab:nae_geo}. 
\par
We now need to construct the other input important to the residual calculation which is,
\begin{equation}
    q_\mathrm{eff}=\frac{1}{\pi}\frac{1}{k_\perp\rho_i}\frac{\omega_d}{\omega_t},
\end{equation}
whose definition is meant to take the place of $q$ in the RH residual. See the main text, Section~\ref{sec:field_survey}, for more details, including its connections to banana widths (roughly $\sim \rho_i q_\mathrm{eff}/\sqrt{\Delta}$) and the transition between the low-mirror and RH residual regimes.
\par
Let us start by finding the amplitude of the drift frequency $\omega_d(\chi)$. The curvature drift is by definition,
\begin{equation}
    \omega_d(\chi)=-v_T\frac{\mathbf{B}\times\nabla B\cdot\nabla\psi}{B^3}\bar{B} k_\psi\rho_i,
\end{equation}
where we have defined the ion Larmor radius $\rho_i=m_iv_T/q_i\bar{B}$ with respect to some reference field $\bar{B}$. The triple vector product may be directly computed using the contravariant Boozer coordinate basis in the near-axis framework \citep[Eq.~(45)]{jorge2020naeturb}\footnote{The expression in \cite{jorge2020naeturb} has an incorrect additional factor of $B_0$, as can be checked dimensionally. This typo is unimportant.}, which yields 
\begin{equation}
    \omega_d(\chi)=-v_T B_0r\eta k_\psi\rho_i \sin\chi+O(r^2).
\end{equation}
The coefficient $\omega_d$ may be directly read-off from the amplitude of this expression. Note here that $\eta$ plays a primary role in controlling the magnitude of the radial drift, as it controls the magnitude of the magnetic field magnitude gradients. 
\par 
To make sense of the typical magnitude of $\eta$, it is convenient to introduce the description of flux surface shapes in the near-axis framework. Flux surfaces are defined as a function of Boozer coordinates with respect to the magnetic axis, $\mathbf{r}_0$, in the Frenet-Serret basis $\{\hat{b},\hat{\kappa},\hat{\tau}\}$ (tangent, normal and binormal) of the latter, so that $\mathbf{r}(\psi,\theta,\varphi)-\mathbf{r}_0=X\hat{\kappa}+Y\hat{\tau}+Z\hat{b}$. Thus $X$ is a function that gives the distance from flux surfaces to the axis along the normal to the latter. To leading order this is proportional to $X_1=r\eta/\kappa$, while along the binormal it scales like $Y_1\sim\kappa/\eta$ \citep[Eq.~(2.13)]{landreman2019}. Thus, in order to avoid extreme shaping $\eta\sim\kappa$ \citep{rodriguez2023constructing}. As $\kappa$ is generally a function of the toroidal angle and $\eta$ is not, the shaping of flux surfaces will change toroidally, but one may take the curvature as a scale for $\eta$. In the case of a circular cross section tokamak one may show that $\eta=1/R$. This relation between the variation of the magnetic field and the curvature of the axis (a field line after all) is a physical consequence of the relation between the bending field lines and magnetic pressure.
\par
We now need to find an expression for the transit time $\omega_t=v_T/L_d$, where $L_d$ is the connection length; the distance from the trough to the top of the well. We thus need to compute $\ell$, the distance along the field line. In quasisymmetry the length is simply a rescaled form of the Boozer toroidal angle $\varphi$, so that \citep[Eq.~(A20)]{landreman2019}
\begin{equation}
    \frac{\mathrm{d}\chi}{\mathrm{d}\ell}\approx\frac{\bar{\iota}}{R_\mathrm{ax}},
\end{equation}
where $R_\mathrm{ax}=L_\mathrm{ax}/2\pi$ and $L_\mathrm{ax}$ is the length of the magnetic axis, and $\bar{\iota}=\iota-N$. Given that in Eq.~(\ref{eqn:nae_qs_B}) the magnetic field has a well of halfwidth $\pi$, then $L_d\approx \pi R_\mathrm{ax}/\bar{\iota}$ and,
\begin{equation}
    \omega_t=\bar{\iota}\frac{v_T}{\pi R_\mathrm{ax}}.
\end{equation}
Finally, let us consider the normalized perpendicular wavenumber $(k_\perp\rho_i)^2=\langle|\nabla\psi|^2\rangle(k_\psi\rho_i)^2$. Note how we are using an averaged form of the flux expansion, which makes the FLR parameter constant, as assumed in our model construction. The particular form of $k_\perp\rho_i$ is motivated by the involvement of $b=(k_\perp\rho_i)^2/2$ in the residual, where it appears flux surface averaged \citep{plunk2024residual} (including variation along the line would be straightforward). We need $|\nabla\psi|^2$ from the near-axis description of the field; using the contravariant basis once again \citep[Eq.~(41)]{jorge2020naeturb}, 
\begin{equation}
    |\nabla\psi|^2\approx r^2\left(B_0\frac{\kappa}{\eta}\right)^2\left[\left(\frac{\eta}{\kappa}\right)^4\sin^2\chi+\left(\cos\chi-\sigma\sin\chi\right)^2\right],
\end{equation}
where $\sigma$ is a function of the toroidal angle $\varphi$, result of solving a non-linear Riccati equation \citep{garrenboozer1991b,landreman2019}. The flux surface average of this expression can be carried out straightforwardly, using to leading order $\langle\dots\rangle\approx\int\mathrm{d}\chi\mathrm{d}\varphi\dots/(4\pi^2)$,
\begin{equation}
    \left\langle|\nabla\psi|^2\right\rangle\approx(rB_0\hat{\mathcal{G}})^2,\label{eq:Ghat}
\end{equation}
where,
\begin{equation}
    \hat{\mathcal{G}}^2=\frac{1}{4\pi}\int_0^{2\pi}\left(\frac{\kappa}{\eta}\right)^2\left(1+\sigma^2+\frac{\eta^4}{\kappa^4}\right)\mathrm{d}\varphi. \label{eqn:G_qs}
\end{equation}
The involvement of $\sigma$ makes this geometric quantity rather obscure. In fact $\sigma$ is directly related to the shaping of flux surfaces as $Y_1=(\kappa/\eta)(\sin\chi+\sigma \cos\chi)$ \citep[Eq.~(2.13)]{landreman2019}, but its interpretation in simple terms is difficult \citep{rodriguez2023mhd}. Although it may be understood roughly as a measure of the rotation of the elliptical cross-sections near the axis respect to the Frenet-Serret frame \citep[Eq.~(B4a)]{rodriguez2023mhd}, it also affects the elongation of flux surfaces. It would be beneficial in the discussion, thus, to provide a more direct geometric interpretation to $\hat{\mathcal{G}}$. We do so using \cite[Eq.~(3.2a)]{rodriguez2023mhd} to write,
\begin{equation}
    \hat{\mathcal{G}}^2=\frac{1}{2\pi}\int_0^{2\pi}\frac{1}{\sin 2e}\mathrm{d}\varphi \label{eqn:G_e}
\end{equation}
where $\mathcal{E}=\tan e$ and $\mathcal{E}$ is the elongation of the flux surfaces in the plane normal to the axis as a function of $\varphi$. The angle $e\in(0,\pi/2)$ may be interpreted as the angle subtended by a right-angle triangle with the major and minor axes of the ellipse as catheti. Thus the geometric factor $\hat{\mathcal{G}}$ is a direct measure of the flux surface elongation. A value of $\hat{\mathcal{G}}=1$ corresponds to all cross-section being circular, any amount of shaping leading to $\hat{\mathcal{G}}>1$. 
\par
Putting everything together into $q_\mathrm{eff}$,
\begin{equation}
    q_\mathrm{eff}=\frac{1}{\iota-N}\frac{\eta R_\mathrm{ax}}{\hat{\mathcal{G}}}.
\end{equation}

\subsubsection{Tokamak limit}
The case of the axisymmetric tokamak is a particularly simple limit of this. Considering the limit of $\kappa\rightarrow1/R$, where $R$ is the major radius, then $R_\mathrm{ax}\rightarrow R$ and all quantities become $\varphi$-independent. Then, we may write $q_\mathrm{eff}=q(\eta R)/\hat{\mathcal{G}}_\mathrm{tok}$, where $q=1/\iota$ is the safety factor and, $\hat{\mathcal{G}}_\mathrm{tok}^2=1/\sin 2e$. If we then consider a circular cross-section tokamak (where $e=\pi/4$), then $\eta=1/R$,  $\hat{\mathcal{G}}=1$, and thus $q_\mathrm{eff}=q$. This is why we have defined $q_\mathrm{eff}$ the way we have. As a reference $\hat{\mathcal{G}}=2$ corresponds to $e=\pi/8$ and thus an elongation $\mathcal{E}\approx0.4$.

\subsection{Quasi-isodynamic fields}
Let us write the magnetic field of an exactly omnigeneous, QI, stellarator-symmetric field near the axis \citep[Eq.~(6.1)]{plunk2019direct} \citep[Eqs.~(8-9a)]{rodriguez2023higher},
\begin{equation}
    B=B_0(\varphi)\left[1-rd(\varphi)\sin\alpha+O(r^2)\right], \label{eqn:nae_B_qi}
\end{equation}
where $B_0(\varphi)$ and $d(\varphi)$ are even and odd functions of $\varphi$ respectively. The latter is required for the fulfilment of omnigeneity. Note that $B$ is here an explicit function of $\alpha$, which unless the rotational transform is integer, makes $B$ a non-periodic function. This is the well-known impossibility of achieving omnigeneity exactly to leading order near the axis with poloidal $|\mathbf{B}|$ contours \citep{plunk2019direct}. Acknowledging that in practice omnigeneity will have to be broken in some \textit{buffer region} near the tops \citep{plunk2019direct,camacho-mata-2022}, we shall consider Eq.~(\ref{eqn:nae_B_qi}) as given.
\par
Let us now consider a simple model for the magnetic field on axis,
\begin{equation}
    B_0(\varphi)=\bar{B}\left(1-\Delta\cos N_\mathrm{nfp}\varphi\right),
\end{equation}
where $\Delta$ is the mirror ratio and $N_\mathrm{nfp}$ is the number of field periods (the toroidal $N_\mathrm{nfp}$-fold symmetry). Unlike in the QS scenario, the control of the on-axis magnetic field in a QI configuration gives complete control of the mirror ratio.
\par
The choice of this form of $B_0$ requires the curvature to have vanishing points at $\varphi=n\pi/N_\mathrm{nfp}$ for $n\in\mathbb{Z}$, and non-vanishing first derivative (often referred to as a \textit{first order zero}). Not doing so would lead to the loss of trapped particles as discussed in detail in \cite{rodriguez2023higher}. As a result, the variation in the field $d(\varphi)$ must also share those zeroes with $\kappa$ to avoid extreme shaping (the leading order shaping is analogous to the QS scenario). For now, let us keep it general and construct the necessary coefficients as we did with the QS case. Starting off the drift, and using \cite[Eq.~(37)]{jorge2020naeturb},
\begin{equation}
    \omega_d(\theta)\approx-rv_T\bar{B}\kappa\left(X_{1c}\sin\theta-X_{1s}\cos\theta\right)k_\psi\rho_i, \label{eqn:wd_nae_qi_I}
\end{equation}
where $X_{1c}$ and $X_{1s}$ are the cosine and sine $\theta$-harmonics of $X_1$ to leading order. Following their definition in terms of $B$ \citep[Eq.~(A22)]{landreman2019}, and using the expression for $B$ in Eq.~(\ref{eqn:nae_B_qi}), for an exactly omnigeneous field,
\begin{subequations}
    \begin{align}
        X_{1c}=&\frac{d}{\kappa}\sin\iota\varphi, \\
        X_{1s}=&-\frac{d}{\kappa}\cos\iota\varphi, \\        
    \end{align}
\end{subequations}
so that Eq.~(\ref{eqn:wd_nae_qi_I}) reduces to,
\begin{equation}
    \omega_d(\varphi)=-rv_T\bar{B}d(\varphi)k_\psi\rho_i\cos \alpha.
\end{equation}
We need the amplitude of this function to feed into $q_\mathrm{eff}$, Of course, generally the shape of this function will not be that of a simple sine as in the QS case. However, we may choose the simple form,
\begin{equation}
    d(\varphi)=\bar{d}\sin (N_\mathrm{nfp}\varphi), \label{eqn:d_nae}
\end{equation}
to give an amplitude $\omega_d\approx rv_T\bar{d}\bar{B}\cos\alpha$. Note a significant difference with respect to the QS case, which is the explicit $\alpha$ dependence. The amplitude of the field varies from field-line to field-line. We have lost the field-line equivalence \citep{boozer1983transport,helander2014theory, rodriguez2020necessary} of quasisymmetry. To treat this difference consistently within the residual treatment we would have to treat more carefully the variation of the field over the surface. However, for a rough estimate of the drift amplitude, let us keep it as is for now.
\par
Let us now consider $|\nabla\psi|^2$ \citep[Eq.~(33)]{jorge2020naeturb},
\begin{equation}
    |\nabla\psi|^2=r^2B_0^2\left[\left(X_{1c}\sin\theta-X_{1s}\cos\theta\right)^2+\left(Y_{1c}\sin\theta-Y_{1s}\cos\theta\right)^2\right],
\end{equation}
where for our ideal omnigeneneous field \citep[Eq.~(A25)]{landreman2019},
\begin{subequations}
    \begin{align}
        Y_{1c}=&\frac{\bar{B}}{B_0}\frac{\kappa}{d}\left(\cos\iota\varphi+\sigma\sin\iota\varphi\right), \\
        Y_{1s}=&-\frac{\bar{B}}{B_0}\frac{\kappa}{d}\left(\sigma\cos\iota\varphi-\sin\iota\varphi\right).
    \end{align}
\end{subequations}
Therefore, 
\begin{equation}
    |\nabla\psi|^2\approx r^2B_0^2\left[\left(\frac{d}{\kappa}\right)^2\cos^2\alpha+\left(\frac{\kappa}{d}\frac{\bar{B}}{B_0}\right)^2\left(\sin\alpha+\sigma\cos\alpha\right)^2\right].
\end{equation}
Assuming $\Delta\ll1$ to simplify the flux surface averages and approximate $B_0\approx\bar{B}$, integrating over $\alpha$ and $\varphi$,
\begin{equation}
    \left\langle|\nabla\psi|\right\rangle \approx \left(r\bar{B}\hat{\mathcal{G}}_\mathrm{QI}\right)^2,\label{eq:Ghat-QI}
\end{equation}
where,
\begin{equation}
    \hat{\mathcal{G}}_\mathrm{QI}^2=\frac{1}{4\pi}\int_0^{2\pi}\left(\frac{\kappa}{d}\right)^2\left(1+\sigma^2+\frac{d^4}{\kappa^4}\right)\mathrm{d}\varphi. \label{eqn:G_qi}
\end{equation}
Note the similarity of this expression to the QS geometric factor Eq.~(\ref{eqn:G_qs}). In fact, Eq.~(\ref{eqn:G_qi}) is exactly equivalent to Eq.~(\ref{eqn:G_e}), the expression in terms of the elongation of flux surfaces in the plane normal to the magnetic axis. 
\par
Finally we compute the connection length, which under the approximation of $\Delta\ll1$ we may write as $L_d\approx \pi r_\mathrm{ax}/N_\mathrm{nfp}$. Putting all together,
\begin{equation}
    q_\mathrm{eff}=\frac{1}{N_\mathrm{nfp}}\frac{\bar{d}R_\mathrm{ax}}{\hat{\mathcal{G}}_\mathrm{QI}}\cos\alpha.
\end{equation}
Note how this parameter changes from field line to field line. The contribution to the total residual can be thought of as a sum over wells, where each of these can be thought of separately, thanks to the condition of omnigeneity. As we move along the field line then, we see different wells, which assuming this to be the only element that changes from well to well, and using 
\begin{equation}
    \lim_{N\rightarrow\infty}\frac{1}{N}\sum_{n=0}^N|\cos(2\pi\iota n)|=\frac{1}{2\pi}\int_0^{2\pi}|\cos\alpha|\mathrm{d}\alpha=\frac{2}{\pi},
\end{equation}
by application of Weyl's lemma \citep[Eq.~(2)]{weyl1916gleichverteilung} for irrational $\iota$, we may construct an effective parameter $q_\mathrm{eff}$,
\begin{equation}
    q_\mathrm{eff}=\frac{1}{N_\mathrm{nfp}}\frac{2}{\pi}\frac{\bar{d}R_\mathrm{ax}}{\hat{\mathcal{G}}}.
\end{equation}
We shall not consider here any more sophisticated approach that deals with these variations more carefully or takes additional differences between wells into account.

\bibliographystyle{jpp}

\bibliography{jpp-instructions}

\begin{thebibliography}{75}
\expandafter\ifx\csname natexlab\endcsname\relax\def\natexlab#1{#1}\fi
\def\au#1{#1} \def\ed#1{#1} \def\yr#1{#1}\def\at#1{#1}\def\jt#1{\textit{#1}} \def\bt#1{#1}\def\bvol#1{\textbf{#1}} \def\vol#1{#1} \def\pg#1{#1} \def\publ#1{#1}\def\arxiv#1{#1}\def\org#1{#1}\def\st#1{\textit{#1}}

\bibitem[Abramowitz \& Stegun(1968)]{abramowitz1968handbook}
{\sc \au{Abramowitz, Milton} \& \au{Stegun, Irene~A}} \yr{1968} {\em Handbook of mathematical functions with formulas, graphs, and mathematical tables\/}, ,  \vol{vol.~55}.  \publ{US Government printing office}.

\bibitem[Alonso {\em et~al.\/}(2017)Alonso, S{\'a}nchez, Calvo, Velasco, McCarthy, Chmyga, Eliseev, Estrada, Kleiber, Krupnik {\em et~al.\/}]{alonso2017observation}
{\sc \au{Alonso, JA}, \au{S{\'a}nchez, E}, \au{Calvo, I}, \au{Velasco, JL}, \au{McCarthy, KJ}, \au{Chmyga, A}, \au{Eliseev, LG}, \au{Estrada, T}, \au{Kleiber, R}, \au{Krupnik, LI} \& \au{others}} \yr{2017}  \at{Observation of oscillatory radial electric field relaxation in a helical plasma}.  \jt{Physical Review Letters}  \bvol{118}~(18),  \pg{185002}.

\bibitem[Austin {\em et~al.\/}(2019)Austin, Marinoni, Walker, Brookman, Degrassie, Hyatt, McKee, Petty, Rhodes, Smith {\em et~al.\/}]{austin2019achievement}
{\sc \au{Austin, Max~E}, \au{Marinoni, A}, \au{Walker, ML}, \au{Brookman, MW}, \au{Degrassie, JS}, \au{Hyatt, AW}, \au{McKee, GR}, \au{Petty, CC}, \au{Rhodes, TL}, \au{Smith, SP} \& \au{others}} \yr{2019}  \at{Achievement of reactor-relevant performance in negative triangularity shape in the diii-d tokamak}.  \jt{Physical review letters}  \bvol{122}~(11),  \pg{115001}.

\bibitem[Barnes {\em et~al.\/}(2019)Barnes, Parra \& Landreman]{barnes2019stella}
{\sc \au{Barnes, Michael}, \au{Parra, Felix~I} \& \au{Landreman, Matt}} \yr{2019}  \at{stella: An operator-split, implicit--explicit $\delta$f-gyrokinetic code for general magnetic field configurations}.  \jt{Journal of Computational Physics}  \bvol{391},  \pg{365--380}.

\bibitem[Beidler {\em et~al.\/}(2021)Beidler, Smith, Alonso, Andreeva, Baldzuhn, Beurskens, Borchardt, Bozhenkov, Brunner, Damm {\em et~al.\/}]{beidler2021demonstration}
{\sc \au{Beidler, CD}, \au{Smith, HM}, \au{Alonso, A}, \au{Andreeva, T}, \au{Baldzuhn, J}, \au{Beurskens, MNA}, \au{Borchardt, Matthias}, \au{Bozhenkov, SA}, \au{Brunner, Kai~Jakob}, \au{Damm, Hannes} \& \au{others}} \yr{2021}  \at{Demonstration of reduced neoclassical energy transport in wendelstein 7-x}.  \jt{Nature}  \bvol{596}~(7871),  \pg{221--226}.

\bibitem[Bender \& Orszag(2013)]{bender2013advanced}
{\sc \au{Bender, Carl~M} \& \au{Orszag, Steven~A}} \yr{2013} {\em Advanced mathematical methods for scientists and engineers I: Asymptotic methods and perturbation theory\/}.  \publ{Springer Science \& Business Media}.

\bibitem[Bernardin {\em et~al.\/}(1986)Bernardin, Moses \& Tataronis]{bernardin1986}
{\sc \au{Bernardin, M.~P.}, \au{Moses, R.~W.} \& \au{Tataronis, J.~A.}} \yr{1986}  \at{Isodynamical (omnigenous) equilibrium in symmetrically confined plasma configurations}.  \jt{The Physics of Fluids}  \bvol{29}~(8),  \pg{2605--2611}.

\bibitem[Boozer(1983{\natexlab{{\em a\/}}})]{boozer1983}
{\sc \au{Boozer, Allen~H.}} \yr{1983{\natexlab{{\em a\/}}}}  \at{Transport and isomorphic equilibria}.  \jt{The Physics of Fluids}  \bvol{26}~(2),  \pg{496--499}.

\bibitem[Boozer(1983{\natexlab{{\em b\/}}})]{boozer1983transport}
{\sc \au{Boozer, Allen~H}} \yr{1983{\natexlab{{\em b\/}}}}  \at{Transport and isomorphic equilibria}.  \jt{The Physics of Fluids}  \bvol{26}~(2),  \pg{496--499}.

\bibitem[Boozer(1998)]{boozer1998stellarator}
{\sc \au{Boozer, Allen~H}} \yr{1998}  \at{What is a stellarator?}  \jt{Physics of Plasmas}  \bvol{5}~(5),  \pg{1647--1655}.

\bibitem[Camacho~Mata {\em et~al.\/}(2022)Camacho~Mata, Plunk \& Jorge]{camacho-mata-2022}
{\sc \au{Camacho~Mata, K.}, \au{Plunk, G.~G.} \& \au{Jorge, R.}} \yr{2022}  \at{Direct construction of stellarator-symmetric quasi-isodynamic magnetic configurations}.  \jt{Journal of Plasma Physics}  \bvol{88}~(5),  \pg{905880503}.

\bibitem[Cary \& Shasharina(1997)]{Cary1997}
{\sc \au{Cary, J.~R.} \& \au{Shasharina, S.~G.}} \yr{1997}  \at{{Omnigenity and quasihelicity in helical plasma confinement systems}}.  \jt{Physics of Plasmas}  \bvol{4}~(9),  \pg{3323--3333},  \arxiv{arXiv: https://pubs.aip.org/aip/pop/article-pdf/4/9/3323/12664528/3323\_1\_online.pdf}.

\bibitem[Catto {\em et~al.\/}(2017)Catto, Parra \& Pusztai]{catto2017electromagnetic}
{\sc \au{Catto, Peter~J}, \au{Parra, Felix~I} \& \au{Pusztai, Istv{\'a}n}} \yr{2017}  \at{Electromagnetic zonal flow residual responses}.  \jt{Journal of Plasma Physics}  \bvol{83}~(4),  \pg{905830402}.

\bibitem[Connor {\em et~al.\/}(1980)Connor, Hastie \& Taylor]{connor1980stability}
{\sc \au{Connor, JW}, \au{Hastie, RJ} \& \au{Taylor, JB}} \yr{1980}  \at{Stability of general plasma equilibria. iii}.  \jt{Plasma Physics}  \bvol{22}~(7),  \pg{757}.

\bibitem[Connor {\em et~al.\/}(1983)Connor, Hastie \& Martin]{connor1983effect}
{\sc \au{Connor, J.~W.}, \au{Hastie, R.~J.} \& \au{Martin, T.~J.}} \yr{1983}  \at{Effect of pressure gradients on the bounce-averaged particle drifts in a tokamak}.  \jt{Nuclear fusion}  \bvol{23}~(12),  \pg{1702}.

\bibitem[Connor {\em et~al.\/}(1978)Connor, Hastie \& Taylor]{CHT}
{\sc \au{Connor, J~W}, \au{Hastie, R~J} \& \au{Taylor, J~B}} \at{ \yr{1978} } \jt{Phys. Rev. Lett.}  \bvol{40}~(6),  \pg{396}.

\bibitem[Conway {\em et~al.\/}(2021)Conway, Smolyakov \& Ido]{conway2021geodesic}
{\sc \au{Conway, Garrard~D}, \au{Smolyakov, Andrei~I} \& \au{Ido, Takeshi}} \yr{2021}  \at{Geodesic acoustic modes in magnetic confinement devices}.  \jt{Nuclear Fusion}  \bvol{62}~(1),  \pg{013001}.

\bibitem[Diamond {\em et~al.\/}(2005)Diamond, Itoh, Itoh \& Hahm]{diamond2005zonal}
{\sc \au{Diamond, Patrick~H}, \au{Itoh, SI}, \au{Itoh, K} \& \au{Hahm, TS}} \yr{2005}  \at{Zonal flows in plasma—a review}.  \jt{Plasma Physics and Controlled Fusion}  \bvol{47}~(5),  \pg{R35}.

\bibitem[Fried \& Conte(2015)]{fried2015plasma}
{\sc \au{Fried, Burton~D} \& \au{Conte, Samuel~D}} \yr{2015} {\em The plasma dispersion function: the Hilbert transform of the Gaussian\/}.  \publ{Academic press}.

\bibitem[Galeev {\em et~al.\/}(1969)Galeev, Sagdeev, Furth \& Rosenbluth]{galeev1969plasma}
{\sc \au{Galeev, Albert~A}, \au{Sagdeev, RZ}, \au{Furth, HP} \& \au{Rosenbluth, MN}} \yr{1969}  \at{Plasma diffusion in a toroidal stellarator}.  \jt{Physical Review Letters}  \bvol{22}~(11),  \pg{511}.

\bibitem[Gao {\em et~al.\/}(2006)Gao, Itoh, Sanuki \& Dong]{gao2006multiple}
{\sc \au{Gao, Zhe}, \au{Itoh, K}, \au{Sanuki, H} \& \au{Dong, JQ}} \yr{2006}  \at{Multiple eigenmodes of geodesic acoustic mode in collisionless plasmas}.  \jt{Physics of plasmas}  \bvol{13}~(10).

\bibitem[Gao {\em et~al.\/}(2008)Gao, Itoh, Sanuki \& Dong]{gao2008eigenmode}
{\sc \au{Gao, Zhe}, \au{Itoh, K}, \au{Sanuki, H} \& \au{Dong, JQ}} \yr{2008}  \at{Eigenmode analysis of geodesic acoustic modes}.  \jt{Physics of Plasmas}  \bvol{15}~(7).

\bibitem[Garren \& Boozer(1991{\natexlab{{\em a\/}}})]{garrenboozer1991b}
{\sc \au{Garren, D.~A.} \& \au{Boozer, A.~H.}} \yr{1991{\natexlab{{\em a\/}}}}  \at{Existence of quasihelically symmetric stellarators}.  \jt{Physics of Fluids B: Plasma Physics}  \bvol{3}~(10),  \pg{2822--2834}.

\bibitem[Garren \& Boozer(1991{\natexlab{{\em b\/}}})]{garrenboozer1991a}
{\sc \au{Garren, D.~A.} \& \au{Boozer, A.~H.}} \yr{1991{\natexlab{{\em b\/}}}}  \at{Magnetic field strength of toroidal plasma equilibria}.  \jt{Physics of Fluids B: Plasma Physics}  \bvol{3}~(10),  \pg{2805--2821}.

\bibitem[Giuliani(2024)]{giuliani2024direct}
{\sc \au{Giuliani, Andrew}} \yr{2024}  \at{Direct stellarator coil design using global optimization: application to a comprehensive exploration of quasi-axisymmetric devices}.  \jt{Journal of Plasma Physics}  \bvol{90}~(3),  \pg{905900303}.

\bibitem[Goodman {\em et~al.\/}(2023)Goodman, Camacho~Mata, Henneberg, Jorge, Landreman, Plunk, Smith, Mackenbach, Beidler, Helander \& et~al.]{goodman2023constructing}
{\sc \au{Goodman, A.G.}, \au{Camacho~Mata, K.}, \au{Henneberg, S.A.}, \au{Jorge, R.}, \au{Landreman, M.}, \au{Plunk, G.G.}, \au{Smith, H.M.}, \au{Mackenbach, R.J.J.}, \au{Beidler, C.D.}, \au{Helander, P.} \& \au{et~al.}} \yr{2023}  \at{Constructing precisely quasi-isodynamic magnetic fields}.  \jt{Journal of Plasma Physics}  \bvol{89}~(5),  \pg{905890504}.

\bibitem[Gradshteyn \& Ryzhik(2014)]{gradshteyn2014table}
{\sc \au{Gradshteyn, Izrail~Solomonovich} \& \au{Ryzhik, Iosif~Moiseevich}} \yr{2014} {\em Table of integrals, series, and products\/}.  \publ{Academic press}.

\bibitem[Hall \& McNamara(1975{\natexlab{{\em a\/}}})]{Hall}
{\sc \au{Hall, L.~S.} \& \au{McNamara, B.}} \yr{1975{\natexlab{{\em a\/}}}}  \at{{Three‐dimensional equilibrium of the anisotropic, finite‐pressure guiding‐center plasma: Theory of the magnetic plasma}}.  \jt{The Physics of Fluids}  \bvol{18}~(5),  \pg{552--565},  \arxiv{arXiv: https://pubs.aip.org/aip/pfl/article-pdf/18/5/552/12317924/552\_1\_online.pdf}.

\bibitem[Hall \& McNamara(1975{\natexlab{{\em b\/}}})]{hall1975}
{\sc \au{Hall, Laurence~S.} \& \au{McNamara, Brendan}} \yr{1975{\natexlab{{\em b\/}}}}  \at{Three‐dimensional equilibrium of the anisotropic, finite‐pressure guiding‐center plasma: Theory of the magnetic plasma}.  \jt{The Physics of Fluids}  \bvol{18}~(5),  \pg{552--565}.

\bibitem[Hazeltine \& Meiss(2003)]{hazeltine2003plasma}
{\sc \au{Hazeltine, Richard~D} \& \au{Meiss, James~D}} \yr{2003} {\em Plasma confinement\/}.  \publ{Courier Corporation}.

\bibitem[Helander(2014)]{helander2014theory}
{\sc \au{Helander, P.}} \yr{2014}  \at{Theory of plasma confinement in non-axisymmetric magnetic fields}.  \jt{Reports on Progress in Physics}  \bvol{77}~(8),  \pg{087001}.

\bibitem[Helander {\em et~al.\/}(2011)Helander, Mishchenko, Kleiber \& Xanthopoulos]{helander2011oscillations}
{\sc \au{Helander, P}, \au{Mishchenko, A}, \au{Kleiber, R} \& \au{Xanthopoulos, P}} \yr{2011}  \at{Oscillations of zonal flows in stellarators}.  \jt{Plasma Physics and Controlled Fusion}  \bvol{53}~(5),  \pg{054006}.

\bibitem[Helander \& Nührenberg(2009)]{Helander_2009}
{\sc \au{Helander, P.} \& \au{Nührenberg, J.}} \yr{2009}  \at{Bootstrap current and neoclassical transport in quasi-isodynamic stellarators}.  \jt{Plasma Physics and Controlled Fusion}  \bvol{51}~(5),  \pg{055004}.

\bibitem[Helander \& Sigmar(2005)]{helander2005collisional}
{\sc \au{Helander, Per} \& \au{Sigmar, Dieter~J}} \yr{2005} {\em Collisional transport in magnetized plasmas\/}, ,  \vol{vol.~4}.  \publ{Cambridge University Press}.

\bibitem[Ho \& Kulsrud(1987)]{ho1987}
{\sc \au{Ho, Darwin~D.‐M.} \& \au{Kulsrud, Russell~M.}} \yr{1987}  \at{Neoclassical transport in stellarators}.  \jt{The Physics of Fluids}  \bvol{30}~(2),  \pg{442--461}.

\bibitem[Jorge \& Landreman(2020)]{jorge2020naeturb}
{\sc \au{Jorge, Rogerio} \& \au{Landreman, Matt}} \yr{2020}  \at{The use of near-axis magnetic fields for stellarator turbulence simulations}.  \jt{Plasma Physics and Controlled Fusion}  \bvol{63}~(1),  \pg{014001}.

\bibitem[Landreman(2022)]{landreman2022mapping}
{\sc \au{Landreman, Matt}} \yr{2022}  \at{Mapping the space of quasisymmetric stellarators using optimized near-axis expansion}.  \jt{Journal of Plasma Physics}  \bvol{88}~(6),  \pg{905880616}.

\bibitem[Landreman \& Catto(2012)]{landreman2012}
{\sc \au{Landreman, M.} \& \au{Catto, P.~J.}} \yr{2012}  \at{Omnigenity as generalized quasisymmetry}.  \jt{Physics of Plasmas}  \bvol{19}~(5),  \pg{056103}.

\bibitem[Landreman \& Paul(2022)]{landreman2022magnetic}
{\sc \au{Landreman, M.} \& \au{Paul, E.}} \yr{2022}  \at{Magnetic fields with precise quasisymmetry for plasma confinement}.  \jt{Physical Review Letters}  \bvol{128}~(3),  \pg{035001}.

\bibitem[Landreman \& Sengupta(2019)]{landreman2019}
{\sc \au{Landreman, M.} \& \au{Sengupta, W.}} \yr{2019}  \at{Constructing stellarators with quasisymmetry to high order}.  \jt{Journal of Plasma Physics}  \bvol{85}~(6),  \pg{815850601}.

\bibitem[Mikhailov {\em et~al.\/}(2002)Mikhailov, Shafranov, Subbotin, Isaev, N\"{u}hrenberg, Zille \& Cooper]{mikhailov2002}
{\sc \au{Mikhailov, M.~I.}, \au{Shafranov, V.~D.}, \au{Subbotin, A.~A.}, \au{Isaev, M.~Y.}, \au{N\"{u}hrenberg, J.}, \au{Zille, R.} \& \au{Cooper, W.~A.}} \at{ \yr{2002} }  \bvol{42}~(11),  \pg{L23--L26}.

\bibitem[Miller {\em et~al.\/}(1998)Miller, Chu, Greene, Lin-Liu \& Waltz]{miller1998noncircular}
{\sc \au{Miller, RL}, \au{Chu, MS}, \au{Greene, JM}, \au{Lin-Liu, YR} \& \au{Waltz, RE}} \yr{1998}  \at{Noncircular, finite aspect ratio, local equilibrium model}.  \jt{Physics of Plasmas}  \bvol{5}~(4),  \pg{973--978}.

\bibitem[Mishchenko {\em et~al.\/}(2008)Mishchenko, Helander \& K{\"o}nies]{mishchenko2008collisionless}
{\sc \au{Mishchenko, Alexey}, \au{Helander, Per} \& \au{K{\"o}nies, Axel}} \yr{2008}  \at{Collisionless dynamics of zonal flows in stellarator geometry}.  \jt{Physics of Plasmas}  \bvol{15}~(7).

\bibitem[Mishchenko \& Kleiber(2012)]{mishchenko2012zonal}
{\sc \au{Mishchenko, Alexey} \& \au{Kleiber, Ralf}} \yr{2012}  \at{Zonal flows in stellarators in an ambient radial electric field}.  \jt{Physics of Plasmas}  \bvol{19}~(7).

\bibitem[Monreal {\em et~al.\/}(2016)Monreal, Calvo, S{\'a}nchez, Parra, Bustos, K{\"o}nies, Kleiber \& G{\"o}rler]{monreal2016residual}
{\sc \au{Monreal, Pedro}, \au{Calvo, Iv{\'a}n}, \au{S{\'a}nchez, Edilberto}, \au{Parra, F{\'e}lix~I}, \au{Bustos, Andr{\'e}s}, \au{K{\"o}nies, Axel}, \au{Kleiber, Ralf} \& \au{G{\"o}rler, Tobias}} \yr{2016}  \at{Residual zonal flows in tokamaks and stellarators at arbitrary wavelengths}.  \jt{Plasma Physics and Controlled Fusion}  \bvol{58}~(4),  \pg{045018}.

\bibitem[Monreal {\em et~al.\/}(2017)Monreal, S{\'a}nchez, Calvo, Bustos, Parra, Mishchenko, K{\"o}nies \& Kleiber]{monreal2017semianalytical}
{\sc \au{Monreal, Pedro}, \au{S{\'a}nchez, Edilberto}, \au{Calvo, Iv{\'a}n}, \au{Bustos, Andr{\'e}s}, \au{Parra, F{\'e}lix~I}, \au{Mishchenko, Alexey}, \au{K{\"o}nies, Axel} \& \au{Kleiber, Ralf}} \yr{2017}  \at{Semianalytical calculation of the zonal-flow oscillation frequency in stellarators}.  \jt{Plasma Physics and Controlled Fusion}  \bvol{59}~(6),  \pg{065005}.

\bibitem[Mukhovatov \& Shafranov(1971)]{mukhovatov1971}
{\sc \au{Mukhovatov, VS} \& \au{Shafranov, VD}} \yr{1971}  \at{Plasma equilibrium in a tokamak}.  \jt{Nuclear Fusion}  \bvol{11}~(6),  \pg{605}.

\bibitem[Mynick(2006)]{mynick2006}
{\sc \au{Mynick, H.~E.}} \yr{2006}  \at{Transport optimization in stellarators}.  \jt{Physics of Plasmas}  \bvol{13}~(5),  \pg{058102}.

\bibitem[Nemov {\em et~al.\/}(1999)Nemov, Kasilov, Kernbichler \& Heyn]{nemov1999}
{\sc \au{Nemov, V.~V.}, \au{Kasilov, S.~V.}, \au{Kernbichler, W.} \& \au{Heyn, M.~F.}} \yr{1999}  \at{Evaluation of $1/\nu$ neoclassical transport in stellarators}.  \jt{Physics of Plasmas}  \bvol{6}~(12),  \pg{4622--4632}.

\bibitem[N{\"u}hrenberg \& Zille(1988)]{nuhren1988}
{\sc \au{N{\"u}hrenberg, J.} \& \au{Zille, R.}} \yr{1988}  \at{Quasi-helically symmetric toroidal stellarators}.  \jt{Physics Letters A}  \bvol{129}~(2),  \pg{113 -- 117}.

\bibitem[Nührenberg(2010)]{Nührenberg_2010}
{\sc \au{Nührenberg, Jürgen}} \yr{2010}  \at{Development of quasi-isodynamic stellarators}.  \jt{Plasma Physics and Controlled Fusion}  \bvol{52}~(12),  \pg{124003}.

\bibitem[Olver {\em et~al.\/}(2020)Olver, Daalhuis, Lozier, Schneider, Boisvert, Clark, B.~R. Mille~and, Cohl \& M.~A.~McClain]{DLMF}
{\sc \au{Olver, F.~W.~J.}, \au{Daalhuis, A.~B.~Olde}, \au{Lozier, D.~W.}, \au{Schneider, B.~I.}, \au{Boisvert, R.~F.}, \au{Clark, C.~W.}, \au{B.~R. Mille~and, B.~V.~Saunders}, \au{Cohl, H.~S.} \& \au{M.~A.~McClain, eds.}} \yr{2020} Nist digital library of mathematical functions. http://dlmf.nist.gov/, Release 1.0.26 of 2020-03-15.

\bibitem[Plunk \& Helander(2024)]{plunk2024residual}
{\sc \au{Plunk, GG} \& \au{Helander, P}} \yr{2024}  \at{The residual flow in well-optimized stellarators}.  \jt{Journal of Plasma Physics}  \bvol{90}~(2),  \pg{905900205}.

\bibitem[Plunk(2024)]{plunk2024-QI}
{\sc \au{Plunk, G.~G.{, \textit{et al}}}} \yr{2024} A geometric approach to constructing quasi-isodynamic fields. In preparation.

\bibitem[Plunk {\em et~al.\/}(2019)Plunk, Landreman \& Helander]{plunk2019direct}
{\sc \au{Plunk, G.~G.}, \au{Landreman, M.} \& \au{Helander, P.}} \yr{2019}  \at{Direct construction of optimized stellarator shapes. part 3. omnigenity near the magnetic axis}.  \jt{Journal of Plasma Physics}  \bvol{85}~(6),  \pg{905850602}.

\bibitem[Rodriguez {\em et~al.\/}(2020)Rodriguez, Helander \& Bhattacharjee]{rodriguez2020necessary}
{\sc \au{Rodriguez, E.}, \au{Helander, P.} \& \au{Bhattacharjee, A.}} \yr{2020}  \at{Necessary and sufficient conditions for quasisymmetry}.  \jt{Physics of Plasmas}  \bvol{27}~(6),  \pg{062501}.

\bibitem[Rodriguez {\em et~al.\/}(2022)Rodriguez, Sengupta \& Bhattacharjee]{rodriguez2022phases}
{\sc \au{Rodriguez, E.}, \au{Sengupta, W.} \& \au{Bhattacharjee, A.}} \yr{2022}  \at{Phases and phase-transitions in quasisymmetric configuration space}.  \jt{Plasma Physics and Controlled Fusion}  \bvol{64}~(10),  \pg{105006}.

\bibitem[Rodr{\'\i}guez {\em et~al.\/}(2023)Rodr{\'\i}guez, Sengupta \& Bhattacharjee]{rodriguez2023constructing}
{\sc \au{Rodr{\'\i}guez, E}, \au{Sengupta, W} \& \au{Bhattacharjee, A}} \yr{2023}  \at{Constructing the space of quasisymmetric stellarators through near-axis expansion}.  \jt{Plasma Physics and Controlled Fusion}  \bvol{65}~(9),  \pg{095004}.

\bibitem[Rodríguez(2023)]{rodriguez2023mhd}
{\sc \au{Rodríguez, E.}} \yr{2023}  \at{Magnetohydrodynamic stability and the effects of shaping: a near-axis view for tokamaks and quasisymmetric stellarators}.  \jt{Journal of Plasma Physics}  \bvol{89}~(2),  \pg{905890211}.

\bibitem[Rodríguez {\em et~al.\/}(2020)Rodríguez, Helander \& Bhattacharjee]{rodriguez2020}
{\sc \au{Rodríguez, E.}, \au{Helander, P.} \& \au{Bhattacharjee, A.}} \yr{2020}  \at{Necessary and sufficient conditions for quasisymmetry}.  \jt{Physics of Plasmas}  \bvol{27}~(6),  \pg{062501}.

\bibitem[Rodríguez \& Plunk(2023)]{rodriguez2023higher}
{\sc \au{Rodríguez, E.} \& \au{Plunk, G.~G.}} \yr{2023}  \at{{Higher order theory of quasi-isodynamicity near the magnetic axis of stellarators}}.  \jt{Physics of Plasmas}  \bvol{30}~(6),  \pg{062507}.

\bibitem[Rosenbluth \& Hinton(1998)]{rosenbluth1998poloidal}
{\sc \au{Rosenbluth, MN} \& \au{Hinton, FL}} \yr{1998}  \at{Poloidal flow driven by ion-temperature-gradient turbulence in tokamaks}.  \jt{Physical review letters}  \bvol{80}~(4),  \pg{724}.

\bibitem[Schiff(2013)]{schiff2013laplace}
{\sc \au{Schiff, Joel~L}} \yr{2013} {\em The Laplace transform: theory and applications\/}.  \publ{Springer Science \& Business Media}.

\bibitem[Skovoroda(2005)]{skovoroda2005}
{\sc \au{Skovoroda, A.~A.}} \yr{2005}  \at{3d toroidal geometry of currentless magnetic configurations with improved confinement}.  \jt{Plasma Physics and Controlled Fusion}  \bvol{47}~(11),  \pg{1911--1924}.

\bibitem[Spitzer~Jr(1958)]{spitzer1958stellarator}
{\sc \au{Spitzer~Jr, Lyman}} \yr{1958}  \at{The stellarator concept}.  \jt{The Physics of Fluids}  \bvol{1}~(4),  \pg{253--264}.

\bibitem[Stringer(1972)]{stringer1972effect}
{\sc \au{Stringer, TE}} \yr{1972}  \at{Effect of the magnetic field ripple on diffusion in tokamaks}.  \jt{Nuclear Fusion}  \bvol{12}~(6),  \pg{689}.

\bibitem[Sugama \& Watanabe(2005)]{sugama2005dynamics}
{\sc \au{Sugama, H} \& \au{Watanabe, T-H}} \yr{2005}  \at{Dynamics of zonal flows in helical systems}.  \jt{Physical review letters}  \bvol{94}~(11),  \pg{115001}.

\bibitem[Sugama \& Watanabe(2006)]{sugama2006collisionless}
{\sc \au{Sugama, Hideo} \& \au{Watanabe, T-H}} \yr{2006}  \at{Collisionless damping of zonal flows in helical systems}.  \jt{Physics of Plasmas}  \bvol{13}~(1).

\bibitem[Takahasi \& Mori(1974)]{takahasi1974double}
{\sc \au{Takahasi, Hidetosi} \& \au{Mori, Masatake}} \yr{1974}  \at{Double exponential formulas for numerical integration}.  \jt{Publications of the Research Institute for Mathematical Sciences}  \bvol{9}~(3),  \pg{721--741}.

\bibitem[Watanabe {\em et~al.\/}(2008)Watanabe, Sugama \& Ferrando-Margalet]{watanabe2008reduction}
{\sc \au{Watanabe, T-H}, \au{Sugama, H} \& \au{Ferrando-Margalet, S}} \yr{2008}  \at{Reduction of turbulent transport with zonal flows enhanced in helical systems}.  \jt{Physical review letters}  \bvol{100}~(19),  \pg{195002}.

\bibitem[Wesson(2011)]{wessonTok}
{\sc \au{Wesson, John}} \yr{2011} {\em {Tokamaks; 4th ed.}\/}. {\em International series of monographs on physics\/} .  \publ{Oxford: Oxford Univ. Press}.

\bibitem[Weyl(1916)]{weyl1916gleichverteilung}
{\sc \au{Weyl, Hermann}} \yr{1916}  \at{{\"U}ber die gleichverteilung von zahlen mod. eins}.  \jt{Mathematische Annalen}  \bvol{77}~(3),  \pg{313--352}.

\bibitem[Xanthopoulos {\em et~al.\/}(2011)Xanthopoulos, Mischchenko, Helander, Sugama \& Watanabe]{xanthopoulos2011zonal}
{\sc \au{Xanthopoulos, P}, \au{Mischchenko, A}, \au{Helander, P}, \au{Sugama, H} \& \au{Watanabe, T-H}} \yr{2011}  \at{Zonal flow dynamics and control of turbulent transport in stellarators}.  \jt{Physical review letters}  \bvol{107}~(24),  \pg{245002}.

\bibitem[Xiao \& Catto(2006)]{xiao2006short}
{\sc \au{Xiao, Yong} \& \au{Catto, Peter~J}} \yr{2006}  \at{Short wavelength effects on the collisionless neoclassical polarization and residual zonal flow level}.  \jt{Physics of Plasmas}  \bvol{13}~(10).

\bibitem[Xiao {\em et~al.\/}(2007)Xiao, Catto \& Dorland]{xiao2007effects}
{\sc \au{Xiao, Yong}, \au{Catto, Peter~J} \& \au{Dorland, William}} \yr{2007}  \at{Effects of finite poloidal gyroradius, shaping, and collisions on the zonal flow residual}.  \jt{Physics of plasmas}  \bvol{14}~(5).

\end{thebibliography}

\end{document}